\documentclass[twoside,american]{elsarticle}
\usepackage[T1]{fontenc}
\usepackage[utf8]{inputenc}
\usepackage{amsmath}
\usepackage{amssymb}
\usepackage{graphicx}

\makeatletter

\journal{Physica A}

\ifdefined\showcaptionsetup
 \PassOptionsToPackage{caption=false}{subfig}
\fi
\usepackage{subfig}
\makeatother

\usepackage{babel}
\begin{document}
\begin{frontmatter}
\title{Exploring the variational method for thermodynamic models}
\author[rvt]{O. Urbański}
\ead{oliwier1459@gmail.com}
\address[rvt]{Faculty of Physics, Adam Mickiewicz University of Poznań, Uniwersytetu
Poznańskiego 2, 61-614 Poznań, Poland}
\begin{abstract}
This work explores the possibilities of the Gibbs-Bogoliubov-Feynman
variational method, aiming at finding room for designing various drawing
schemes. For example, mean-field approximation can be viewed as a
result of using site-independent drawing in the variational method.
In subsequent sections, progressively complex drawing procedures are
presented, starting from site-independent drawing in the $k$-space.
In the next, each site in the real-space is again drawn independently,
which is followed by an adjustable linear transformation $T$. Both
approaches are presented on the discrete Ginzburg-Landau model. Subsequently,
a percolation-based procedure for the Ising model is developed. It
shows a general way of handling multi-stage drawing schemes. Critical
inverse temperatures are obtained in two and three dimensions with
a few percent discrepancy from exact values. Finally, it is shown
that results in the style of the real-space renormalization group
can be achieved by suitable fractal-like drawing. This facilitates
a new straight-forward approach to establishing the renormalization
transformation, but primarily provides a new view on the method. While
the first two approaches are capable of capturing long-range correlations,
they are not able to reproduce the critical behavior accurately. The
main findings of the paper are developing the method of handling intricate
drawing procedures and identifying the need of fractality in these
schemes to grasp the critical behavior.
\end{abstract}
\begin{keyword}
thermal variational method \sep critical phenomena \sep many-body
complexity \sep mean-field approximation \sep renormalization group
\sep drawing of states
\end{keyword}
\end{frontmatter}

\section{Introduction}

Thermodynamic models of statistical physics, especially those exhibiting
phase transitions, constitute a large family of interesting problems,
which are generally hard (or even impossible) to solve exactly \citep{Binney}.
This is probably because the class of all ``closed-form'' expressions
is extremely small when compared to the class of all possible functions
(for example describing the free energy). If one was able to write
down and effectively understand only polynomials, many others elementary
functions would be impossible to write in such way. However, it is
possible to perform polynomial fitting, often, with a very high accuracy.
Therefore, difficulties in solving a thermodynamic model exactly do
not preclude analytical solutions which are satisfyingly precise.
A natural task which arises from these considerations is to develop
a systematic procedure providing the most accurate results, which
still can be obtained analytically. This is basically what variational
methods attempt to do.

Probably the most popular version of variational methods in physics
refers to finding ground states of quantum systems. This is described,
among numerous other textbooks, by Griffiths \citep{Griffiths}. The
Gibbs-Bogoliubov-Feynman variational principle is designed to deal
with thermodynamic models of classical statistical mechanics, as presented
by Binney et al. and Wang \citep{Binney,Wang}. Its quantum mechanical
generalization is proved by Falk \citep{Falk}.

The Gibbs-Bogoliubov-Feynman method consists in using some trial Hamiltonian
containing adjustable parameters, which can be exactly solved, to
minimize free energy. Thinking more from the side of probability theory,
a scheme for drawing states has to be proposed, for which both the
mean energy and entropy can be calculated. It does not sound like
a big constraint, so there should be plenty of room to construct creative
drawing schemes, which will better and better approximate thermodynamics
of considered models. This feeling is the main motivation for the
presented work. However, while the mean energy is usually easily determined
for schemes the author came up with, the entropy is a real problem.
If $\Omega$ is a set of all microstates of some classical model and
$p_{\sigma}$ is a probability of drawing state $\sigma\in\Omega$,
then the entropy $S$ is given by $-\sum_{\sigma\in\Omega}p_{\sigma}\ln p_{\sigma}$.
Due to the presence of the logarithm, only factorable $p_{\sigma}$
are easy to handle. Such drawing schemes correspond to drawing independently
some features of the system. In particular, drawing independently
state of every site reproduces mean-field approximation. In fact,
\citep{Binney} introduces the variational method to justify the mean-field
approximation for the Ising model. This correspondence (for any quantum
model) is investigated in \citep{Urba=000144ski} from different perspectives.
Final considerations from there constitute a direct trigger for this
work. The rest of the paper is devoted to surpassing mean-field approximation
in the spirit of the variational method and the idea of drawing states.

In section \ref{sec:Mean-field-in-the} state of every site in the
reciprocal space is drawn independently, which leads to a ``$k$-space
mean-field''. Then it is compared to the standard real-space mean-field.
In section \ref{sec:Including-correlations-as} state of every real-space
site is drawn independently, which is followed by some adjustable
linear transformation $T$. Such a trick is equivalent to including
correlations as adjustable parameters. Section \ref{sec:Percolation-based-drawing-for}
presents a more creative drawing scheme, which is based on the percolation
model. It introduces a general method of handling complex multi-stage
schemes. Finally, section \ref{sec:Fractal-like-drawing-reproducing}
introduces a recursive fractal-like drawing procedure, which reproduces
results in the style of renormalization group.

\section{Formalism of the variational method}

The variational method, in the sense used in this paper, is based
on inequality (6.26) from \citep{Binney}:

\begin{equation}
\mathcal{F}_{\mathrm{true}}\leq\left\langle \mathcal{H}\right\rangle _{0}-\frac{1}{\beta}S_{0},\label{var ineq}
\end{equation}
where $\mathcal{F}_{\mathrm{true}}$ is the true free energy of a
system governed by a Hamiltonian $\mathcal{H}$. $\left\langle \cdots\right\rangle _{0}$
denotes averaging over a trial probability distribution $P_{0}$ of
(classical) configurations. $S_{0}$ is its corresponding entropy
and $\beta$ is the inverse temperature.

Since evaluating average energy and entropy (needed for obtaining
$\mathcal{F}_{\mathrm{true}}$ exactly) using the true probability
distribution given by $e^{-\beta\mathcal{H}}/\mathrm{Tr}\left[e^{-\beta\mathcal{H}}\right]$
is difficult, a tractable $P_{0}$ is used. Then, according to \eqref{var ineq},
the free energy gets overestimated, so such $P_{0}$ is sought so
as to minimize the output.

A drawing scheme (or procedure, with both terms used interchangeably)
is a random process of generating (i. e. choosing) a specific configuration
of the system. Each drawing scheme uniquely produces certain probability
distribution $P_{0}$ and thus associated variational free energy.
Minimizing it provides its best estimate within the considered class
of drawing procedures.

\section{Mean-field in the reciprocal space\label{sec:Mean-field-in-the}}

Mean-field arises from drawing state of each site independently, so
it can be easily extended by partitioning sites into finite clusters
and drawing their states independently. Improvement brought by such
clustering can be checked to grow very slowly with cluster size. This
is because such clusters are still unable to account for long-range
correlations, which are a relevant part of phase transitions \citep{Binney}.
It is a good occasion to mention an interesting procedure developed
by Ferreira et al. \citep{Ferreira}, based on cluster variational
method. It improves the latter by including different cluster types
and extrapolating formulas for the free energy to a number of clusters
exceeding their maximum number fitting geometrically into the lattice.
Nevertheless, it is not a fully variational method in the sense of
this paper and will not be analyzed here.

There is, however, an easy way to draw state of each site independently
and bypass the mentioned issues, provided that these are reciprocal
space sites. Those corresponding to low wave vector values $k$ encode
long-range correlations. Such practice is not different from the mean-field
scheme performed in the $k$-space. This leads to a well-known Gaussian
approximation, which among many other places, is used for example
in \citep{Sengupta}. It is intimately connected with the spherical
model \citep{Binney}, which becomes site-decoupled in the reciprocal
space. Here, the described procedure is presented on the discrete
version of the Ginzburg-Landau model to show, that the variational
method viewpoint allows to think about mean-field in a broader way.

We start with the following Hamiltonian (without loss of generality
the inverse temperature $\beta=1$ throughout this section):

\begin{equation}
\mathcal{H}=-\epsilon\sum_{\left\langle ij\right\rangle }\phi_{i}\phi_{j}+a\sum_{i}\left|\phi_{i}\right|^{2}+b\sum_{i}\left|\phi_{i}\right|^{4},\label{H og}
\end{equation}
where $\epsilon,a,b$ are real parameters ($\epsilon,b>0$), $\phi_{i}$
(a real number) is the local magnetization at site $i$ and $\sum_{\left\langle ij\right\rangle }$
denotes summation over all ordered neighboring pairs of sites $\left(i,j\right)$.
Transition into the $k$-space is performed by means of the discrete
Fourier transform ($N$ is the total number of sites):

\begin{equation}
\phi_{k}=\frac{1}{\sqrt{N}}\sum_{i}\phi_{i}e^{-\mathrm{i}k\cdot i},
\end{equation}
so $\phi_{k}$ are complex numbers constrained by $\phi_{k}=\phi_{-k}^{*}$.
Rewriting the Hamiltonian in terms of $\phi_{k}$ gives:

\begin{align}
\mathcal{H} & =\sum_{k}\left(-\epsilon J_{k}+a\right)\left|\phi_{k}\right|^{2}\nonumber \\
 & +\frac{b}{N}\sum_{k_{1}\cdots k_{4}}\phi_{k_{1}}\phi_{k_{2}}\phi_{k_{3}}\phi_{k_{4}}\delta_{k_{1}+k_{2}+k_{3}+k_{4},0},\label{H}
\end{align}
where $J_{k}=2\sum_{l=1}^{d}\cos k_{l}$ is the lattice dispersion
relation (here evaluated for a cubic $d$-dimensional lattice).

In the supercritical regime, we have $\left\langle \phi_{i}\right\rangle =0$,
so also $\left\langle \phi_{k}\right\rangle =0$. If we assume that
the phases of $\phi_{k}$ are totally random, we also have $\left\langle \phi_{k}^{n}\right\rangle =0$,
for any natural $n\geq1$. This significantly simplifies the mean-field
procedure.

Without loss of generality, it can be assumed that all reciprocal
space sites can be paired according to the rule $k\leftrightarrow-k$,
with only one site remaining without a pair, namely $k=0$. Only one
variable from $\left(\phi_{k},\phi_{-k}\right)$ corresponding to
one pair is considered as independent, because of relation $\phi_{k}=\phi_{-k}^{*}$.
$\phi_{k=0}$ is independent, but necessarily real. Inequality $k>0$
is meant to denote that $\phi_{k}$ belongs to the set of (arbitrarily)
chosen independent variables with $k\neq0$. The mean-field Hamiltonian
governing statistics of $\phi_{k}$ ($k>0$ or $k=0$) is found (as
explained in \citep{Urba=000144ski}) by looking at $\mathcal{H}$
from Eq. \eqref{H} and replacing all variables associated with different
(from $k$ or $-k$) sites by their averages. This scheme leads to
(up to an additive constant and for $k>0$):

\begin{align}
\mathcal{H}_{\mathrm{mf}}\left(k\right) & =2\left(-\epsilon J_{k}+a+\frac{6b}{N}\sum_{q\neq\pm k}\left\langle \left|\phi_{q}\right|^{2}\right\rangle \right)\left|\phi_{k}\right|^{2}\nonumber \\
 & +\frac{6b}{N}\left|\phi_{k}\right|^{4}.\label{Hmf(k)}
\end{align}

In the thermodynamic limit ($N\rightarrow\infty$), the quartic term
in $\phi$ disappears and $\frac{1}{N}\sum_{q\neq\pm k}\rightarrow\left(2\pi\right)^{-d}\int\mathrm{d}^{d}q$.
Writing the self-consistency condition for $\left\langle \left|\phi_{k}\right|^{2}\right\rangle $,
we get:

\begin{equation}
\left\langle \left|\phi_{k}\right|^{2}\right\rangle =\frac{1}{2\left(-\epsilon J_{k}+a+\frac{6b}{\left(2\pi\right)^{d}}\int\mathrm{d}^{d}q\left\langle \left|\phi_{q}\right|^{2}\right\rangle \right)}.\label{eq}
\end{equation}

Let $x=\int\mathrm{d}^{d}q\left\langle \left|\phi_{q}\right|^{2}\right\rangle $.
Integrating Eq. \eqref{eq} over the $k$-space yields:

\begin{equation}
x=\int\mathrm{d}^{d}k\frac{1}{2\left(-\epsilon J_{k}+a+\frac{6b}{\left(2\pi\right)^{d}}x\right)}.\label{eq2}
\end{equation}

Let

\begin{equation}
\mathcal{I}\left(\zeta\right)=\frac{1}{2}\int\mathrm{d}^{d}k\frac{1}{-\epsilon J_{k}+\zeta},
\end{equation}
so that Eq. \eqref{eq2} can be written as:

\begin{equation}
x=\mathcal{I}\left(a+\frac{6b}{\left(2\pi\right)^{d}}x\right)
\end{equation}
or, equivalently:

\begin{equation}
\frac{\left(2\pi\right)^{d}}{6b}\left(\zeta-a\right)=\mathcal{I}\left(\zeta\right).\label{eq3}
\end{equation}

Integral $\mathcal{I}\left(\zeta\right)$ is defined for $\zeta>2d\epsilon$.
For $d\leq2$, it is divergent as $\zeta\rightarrow2d\epsilon^{+}$.
Then, Eq. \eqref{eq3} has solutions for any $a$. This situation
changes for $d>2$, when $\mathcal{I}\left(\zeta\right)$ tends to
a finite limit as $\zeta\rightarrow2d\epsilon^{+}$, which we denote
simply by $\mathcal{I}\left(2d\epsilon\right)$. Then, Eq. \eqref{eq3}
has no solutions for $a<a_{c}$, where $a_{c}$ is given by:

\begin{equation}
a_{c}=2d\epsilon-\frac{6b}{\left(2\pi\right)^{d}}\mathcal{I}\left(2d\epsilon\right).\label{ac}
\end{equation}

This sudden disappearance of a solution to the self-consistency equations,
which were developed only for the supercritical conditions, is a manifestation
of a phase transition. Approaching the critical point from the subcritical
side is also possible within the presented method, but more complicated.

It is natural to compare the mean-field method performed in the reciprocal
space and that in the real space. In the latter, the single-site Hamiltonian
(as opposed to Eq. \eqref{Hmf(k)}) is the same for every site and
reads ($z=2d$ is the coordination number):

\begin{equation}
\mathcal{H}_{\mathrm{mf}}=-z\epsilon\left\langle \phi\right\rangle \phi+a\phi^{2}+b\phi^{4}.
\end{equation}

The self-consistency condition becomes:

\begin{equation}
\left\langle \phi\right\rangle =\frac{\int_{-\infty}^{\infty}\mathrm{d}\phi\,e^{z\epsilon\left\langle \phi\right\rangle \phi-a\phi^{2}-b\phi^{4}}\phi}{\int_{-\infty}^{\infty}\mathrm{d}\phi\,e^{z\epsilon\left\langle \phi\right\rangle \phi-a\phi^{2}-b\phi^{4}}}.\label{self-consistency}
\end{equation}

Mean magnetization $\left\langle \phi\right\rangle $ is therefore
found by localizing an intersection of a straight line (representing
the left-hand-side of Eq. \eqref{self-consistency}) with some curve
(representing the right-hand-side of Eq. \eqref{self-consistency}).
A nonzero solution appears when the slope of that curve at $\left\langle \phi\right\rangle =0$
is greater than $1$. Thus criticality corresponds to:

\begin{equation}
\left[\frac{\partial}{\partial\left\langle \phi\right\rangle }\frac{\int_{-\infty}^{\infty}\mathrm{d}\phi\,e^{z\epsilon\left\langle \phi\right\rangle \phi-a\phi^{2}-b\phi^{4}}\phi}{\int_{-\infty}^{\infty}\mathrm{d}\phi\,e^{z\epsilon\left\langle \phi\right\rangle \phi-a\phi^{2}-b\phi^{4}}}\right]_{\left\langle \phi\right\rangle =0}=1.
\end{equation}

This leads to an equation for the critical value of $\epsilon$, for
given $a,b$:

\begin{equation}
\epsilon_{c}=\frac{1}{z}\frac{\int_{-\infty}^{\infty}\mathrm{d}\phi\,e^{-a\phi^{2}-b\phi^{4}}}{\int_{-\infty}^{\infty}\mathrm{d}\phi\,e^{-a\phi^{2}-b\phi^{4}}\phi^{2}}.\label{ec}
\end{equation}

Now, Eqs. \eqref{ac} and \eqref{ec} can be qualitatively compared.
First of all, the $k$-space mean-field prediction recognizes the
concept of lower critical dimension, but its value is $3$ instead
of the correct value $2$ \citep{Binney}. In the real space mean-field
transition occurs even in $d=1$. Moreover, Eq. \eqref{ac} depends
on the lattice type (trough $J_{k}$ hidden in $\mathcal{I}$), while
Eq. \eqref{ec} refers only to $z$. The most relevant feature of
the $k$-space mean-field is that taking the thermodynamic limit is
vital for identifying the phase transition. However, this is not the
case for the real space mean-field, which thus erroneously suggests
phase transitions in finite systems.

\begin{figure}
\begin{centering}
\includegraphics[scale=0.35]{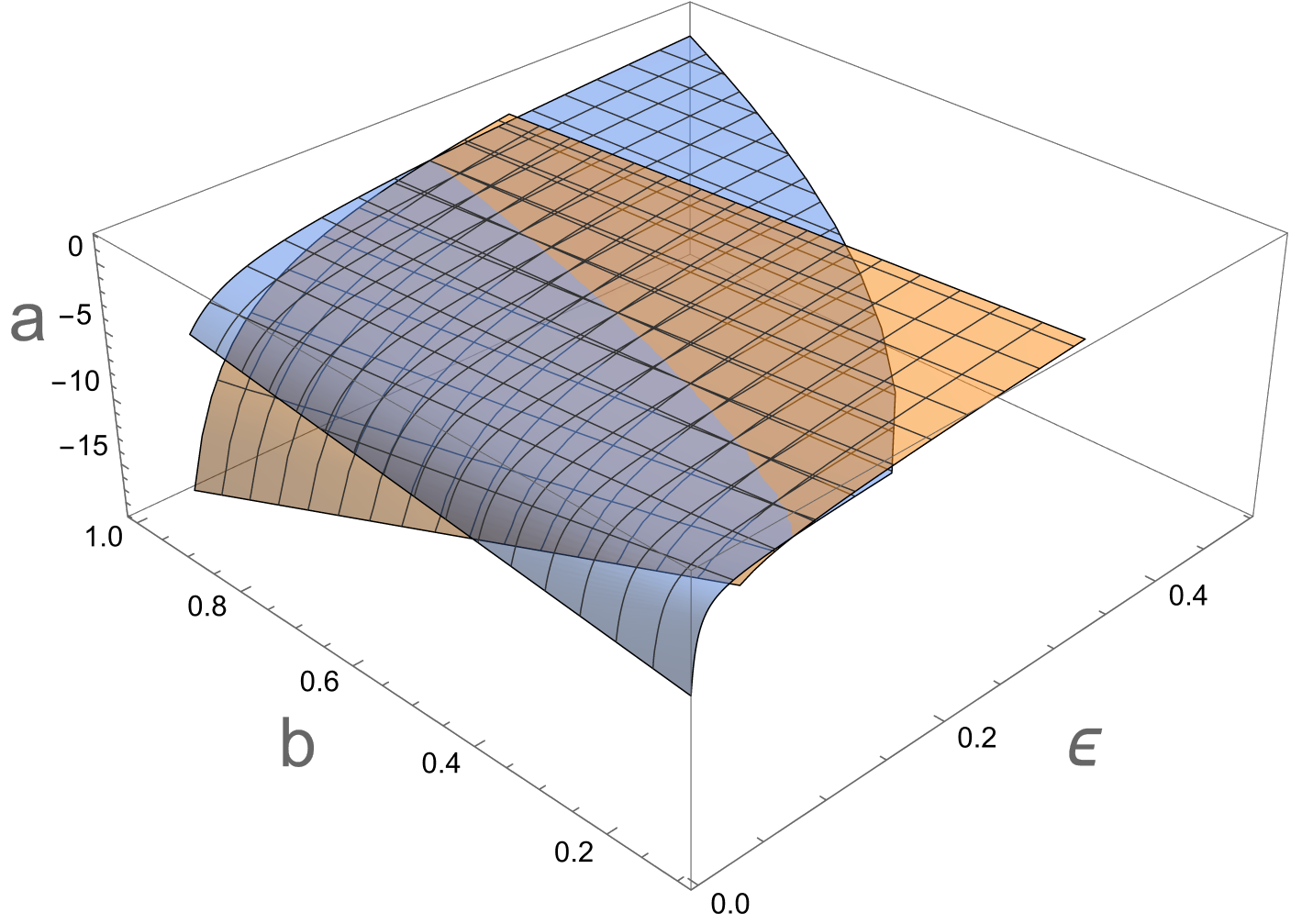}
\par\end{centering}
\centering{}\caption{Critical surfaces comparison between the $k$-space (orange surface
lying at the bottom for low $\epsilon$) and real space (blue surface)
mean-fields for $d=3$.\label{fig:Critical-curves-comparison}}
\end{figure}

Figure \ref{fig:Critical-curves-comparison} shows a quantitative
comparison between the two mean-fields through a plot of their critical
surfaces. The real-space mean-field is known for predicting the transition
too quickly when it is approached from the disordered state. This
is due to a lack of addressing correlations \citep{Binney}. The $k$-space
mean-field improves this issue, but only for not too high $\epsilon$.
Although it involves nonzero correlations of the real-space lattice
sites, their nature is shaped by the requirement that the $k$-space
sites are uncorrelated. Probably, a significant improvement in the
variational method could be achieved if the correlations (in the real-space)
could be variational parameters themselves, allowing for capturing
reliably their spatial structure. This is exactly the task of the
following section.

\section{Including correlations as variable parameters\label{sec:Including-correlations-as}}

Correlations between different sites can be achieved by drawing independently
state of every subsystem. The key is that these subsystems should
not be individual sites, but their ``superpositions''. By this metaphor
is meant the following. Let $\phi$ be a vector composed of $\phi_{i}$
values for every real-space site $i$. We introduce a vector $u$
of $N$ real random variates. Each is drawn independently with some
probability distribution. Drawing $\phi$ is realized by first drawing
$u$ and using relation:

\begin{equation}
\phi=Tu,\label{phi=00003DTu}
\end{equation}
where $T$ is some matrix. Again, the method will be developed for
approaching the critical point from the disordered state, but with
some additional effort, it should be generalizable to the ordered
state as well. Assuming $T$ is invertible, we can write:

\begin{equation}
u_{i}=\sum_{j}\left(T^{-1}\right)_{ij}\phi_{j}.
\end{equation}

Since in the disordered state $\left\langle \phi_{j}\right\rangle =0$,
then $\left\langle u_{i}\right\rangle =0$ for every $i$. Additionally,
a unit variance for each component of $u$ can be assumed, namely
$\left\langle u_{i}^{2}\right\rangle =1$. This is because (provided
that $\left\langle u_{i}^{2}\right\rangle \neq0$), any factor residing
in $u$ can be incorporated into $T$. Thus $\left\langle u_{i}u_{j}\right\rangle =\delta_{ij}$.
Now:

\begin{align}
\left\langle \phi_{i}\phi_{j}\right\rangle  & =\left\langle \left(\sum_{s_{1}}T_{is_{1}}u_{s_{1}}\right)\left(\sum_{s_{2}}T_{js_{2}}u_{s_{2}}\right)\right\rangle \nonumber \\
 & =\sum_{s_{1}s_{2}}T_{is_{1}}T_{js_{2}}\left\langle u_{s_{1}}u_{s_{2}}\right\rangle \nonumber \\
 & =\sum_{s}T_{is}T_{js}=\left[TT^{\mathrm{T}}\right]_{ij}.
\end{align}

$\left\langle \phi_{i}\phi_{j}\right\rangle $ depends only on $i-j$,
so the same is demanded from $\left[TT^{\mathrm{T}}\right]_{ij}$
(a superscript $\mathrm{T}$ denotes transposition). This means, that
after performing a discrete Fourier transform, $TT^{\mathrm{T}}$
becomes diagonal. It can be achieved by constraining to diagonal $T$
in the $k$-space representation (or, equivalently, a translationally
invariant $T_{ij}$). However, $T$ need not to be diagonal, to make
$TT^{\mathrm{T}}$ diagonal. The $k$-space mean-field is an example
of the presented drawing scheme, for specific (and orthogonal) $T$.
To obtain the task of this section, it is sufficient to consider only
translationally invariant $T_{ij}$, which is assumed from now. Symmetry
(i. e. $T_{ij}=T_{ji}$) can also be postulated, since $T_{ij}=T_{i-j,0}$
should equal $T_{j-i,0}=T_{ji}$ due to the point reflection symmetry
of the lattice. Let $T_{k}$ be the diagonal terms in the $k$-space
representation of $T$:

\begin{equation}
T_{k}=\sum_{i}T_{i,0}e^{-\mathrm{i}k\cdot i}.
\end{equation}
Then:

\begin{equation}
\left\langle \phi_{i}\phi_{j}\right\rangle =\frac{1}{N}\sum_{k}T_{k}^{2}e^{\mathrm{i}k\cdot\left(i-j\right)}.\label{corr and Tk}
\end{equation}
Varying $T_{k}$, any spatial structure of the two-point correlation
function can be obtained. It remains to express the variational free
energy in terms of $T_{k}$.

Let $\mathcal{D}\phi\equiv\mathrm{d}\phi_{1}\cdots\mathrm{d}\phi_{N}$
and $\mathcal{D}u\equiv\mathrm{d}u_{1}\cdots\mathrm{d}u_{N}$ be infinitesimal
volumes appearing in integration over the entire phase-space. $T$
is a Jacobian matrix for a $\phi\rightarrow u$ variable change, so
$\mathcal{D}\phi=\det T\,\mathcal{D}u$. Probability density functions
for $\phi$ and $u$ are thus related by $\rho\left(\phi\right)=\tilde{\rho}\left(u\right)\det T^{-1}$.
Let $\mathcal{H}\left(\phi\right)$ be the original Hamiltonian given
by Eq. \eqref{H og} and $\mathcal{H}\left(u\right)$ its form in
the new $u$ variables. Therefore, the free energy of distribution
$\rho\left(\phi\right)$:

\begin{equation}
\mathcal{F}=\int\mathcal{D}\phi\,\rho\left(\phi\right)\mathcal{H}\left(\phi\right)+\frac{1}{\beta}\int\mathcal{D}\phi\,\rho\left(\phi\right)\ln\rho\left(\phi\right)
\end{equation}
can be rewritten as:
\begin{align}
\mathcal{F} & =\int\mathcal{D}u\,\tilde{\rho}\left(u\right)\tilde{\mathcal{H}}\left(u\right)\nonumber \\
 & +\frac{1}{\beta}\int\mathcal{D}u\,\tilde{\rho}\left(u\right)\ln\tilde{\rho}\left(u\right)-\frac{1}{\beta}\mathrm{Tr}\ln T.
\end{align}

Distribution $\tilde{\rho}$, as suggested at the beginning, is factorable:

\begin{equation}
\tilde{\rho}\left(u\right)=\prod_{i}\varrho\left(u_{i}\right),
\end{equation}
which allows to perform mean-field (over $u$). Using the full mean-field
prescription (in which a product $u_{i}u_{j}$ for $i\neq j$ is changed
into $u_{i}\left\langle u_{j}\right\rangle +\left\langle u_{i}\right\rangle u_{j}-\left\langle u_{i}\right\rangle \left\langle u_{j}\right\rangle $
and so on) produces a reduced Hamiltonian $\mathcal{H}_{\mathrm{mf}}$.
The main optimization problem turns into finding an extremum of:

\begin{equation}
\mathcal{F}=-\frac{1}{\beta}\ln\left(\int\mathcal{D}u\,e^{-\beta\mathcal{H}_{\mathrm{mf}}}\right)-\frac{1}{\beta}\mathrm{Tr}\ln T,
\end{equation}
with:

\begin{align}
 & \mathcal{H}_{\mathrm{mf}}=bx_{4}\sum_{j}u_{j}^{4}+\sum_{j}\left\{ 6b\left\langle u^{2}\right\rangle \left(x_{1}^{2}-x_{4}\right)\right.\nonumber \\
 & \left.+\left[T^{\mathrm{T}}\left(-\epsilon J+aI\right)T\right]_{jj}\right\} u_{j}^{2}+3Nb\left\langle u^{2}\right\rangle ^{2}\left(x_{4}-x_{1}^{2}\right)\label{another Hmf}
\end{align}
and ($x_{2}$ and $x_{3}$ will be useful later):

\begin{equation}
x_{1}=\frac{1}{N}\sum_{k}T_{k}^{2}=\sum_{i}T_{i,0}^{2},
\end{equation}
\begin{equation}
x_{2}=\frac{1}{N}\sum_{k}J_{k}T_{k}^{2},
\end{equation}
\begin{equation}
x_{3}=\frac{1}{N}\sum_{k}\ln T_{k},
\end{equation}
\begin{equation}
x_{4}=\sum_{i}T_{i,0}^{4}.
\end{equation}

The $k$-space representation of $T$ can be used to evaluate efficiently
$\left[T^{\mathrm{T}}\left(-\epsilon J+aI\right)T\right]_{jj}=\frac{1}{N}\mathrm{Tr}\left[\left(-\epsilon J+aI\right)T^{2}\right]$
and $\mathrm{Tr}\ln T$. Additionally, we set $\left\langle u^{2}\right\rangle =1$,
$\beta=1$ and decompose $\mathcal{H}_{\mathrm{mf}}$ into a sum of
identical single-site Hamiltonians $\mathcal{H}_{\mathrm{ss}}=Bu^{4}+Au^{2}$
plus some constant. This leads to:

\begin{align}
\frac{\mathcal{F}}{N} & =-\ln\left(\int_{-\infty}^{\infty}\mathrm{d}u\,e^{-\left(Bu^{4}+Au^{2}\right)}\right)\nonumber \\
 & +3b\left(x_{4}-x_{1}^{2}\right)-x_{3},\label{F/N(T)}
\end{align}
where:

\begin{equation}
A=6b\left(x_{1}^{2}-x_{4}\right)-\epsilon x_{2}+ax_{1},
\end{equation}
\begin{equation}
B=bx_{4}.
\end{equation}

Now, we are faced with a purely variational problem, in which correlations
can be directly determined. Specific forms involving a few variable
parameters can be assumed for $T$ or it can be found using numerical
methods. The second option has been used to investigate qualitative
features of the developed method.

\begin{figure*}
\centering{}\subfloat[$T_{k}$]{\centering{}\includegraphics[scale=0.29]{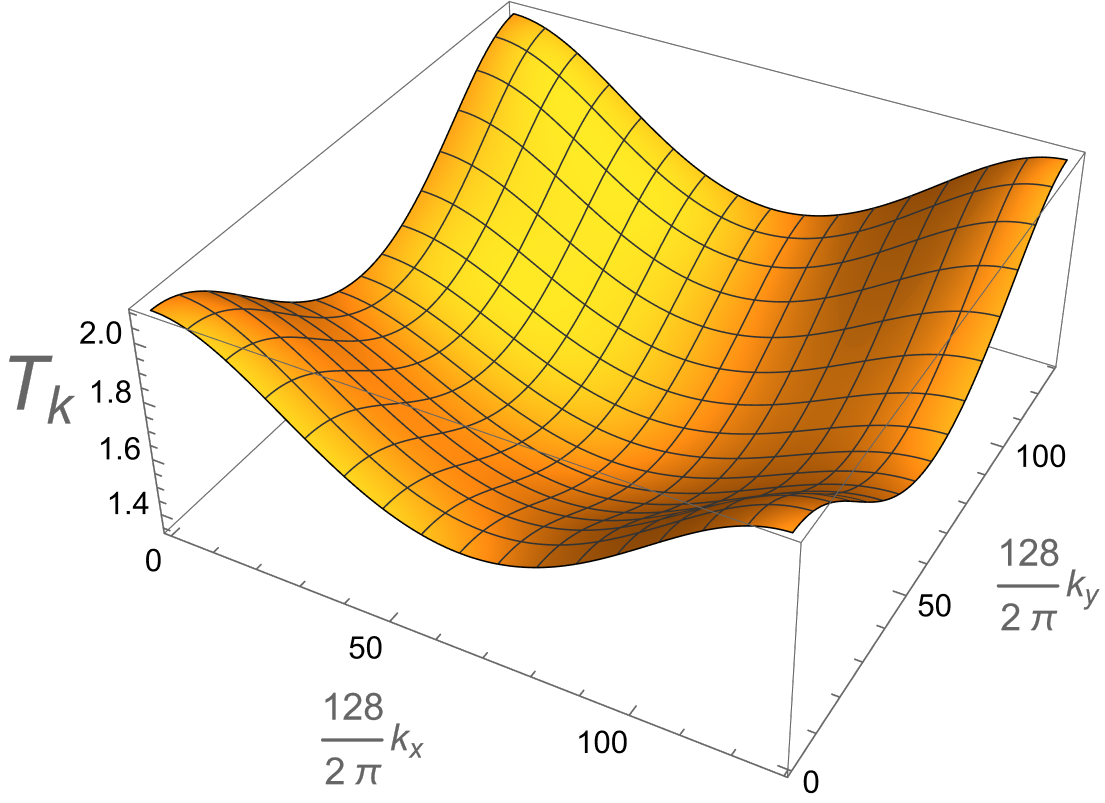}}\hspace{0.2cm}\subfloat[$\ln\left\langle \phi_{i}\phi_{0}\right\rangle $]{\centering{}\includegraphics[scale=0.33]{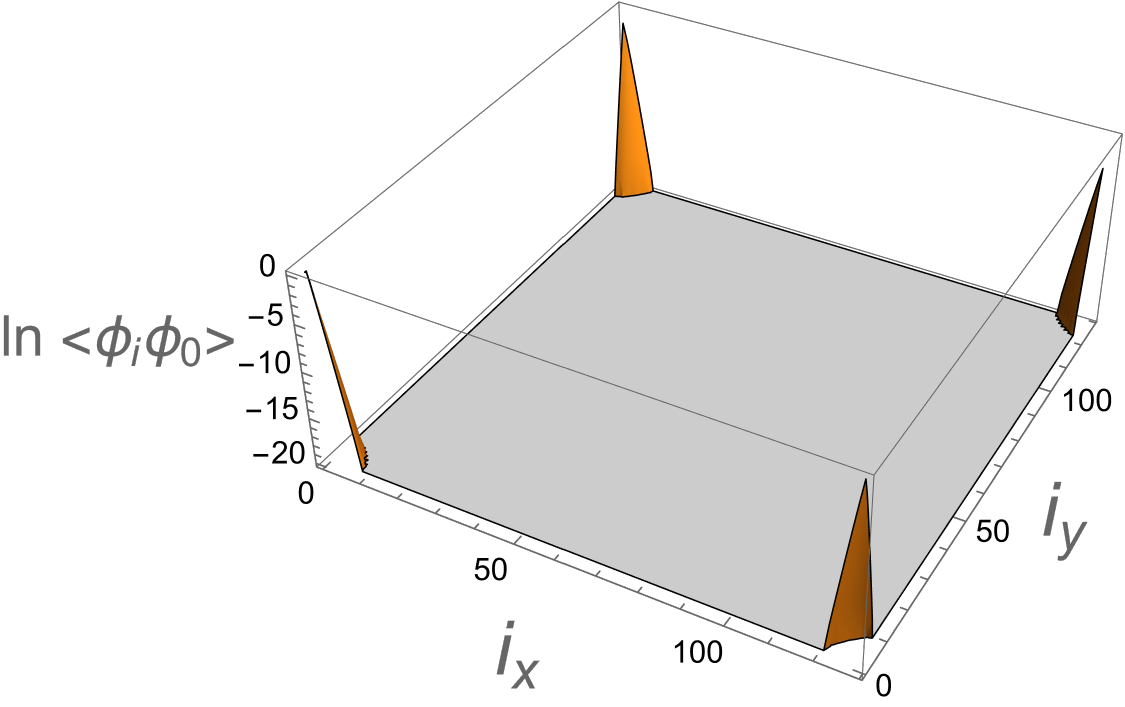}}\caption{Optimal $T_{k}$ and resulting real-space correlation $\left\langle \phi_{i}\phi_{0}\right\rangle $
(on log scale) for exemplary parameters $\left(a,b,\epsilon\right)=\left(-3,0.5,0.3\right)$.\label{fig:Optimal--and}}
\end{figure*}

\begin{figure*}
\centering{}\subfloat[$T_{k}$]{\centering{}\includegraphics[scale=0.29]{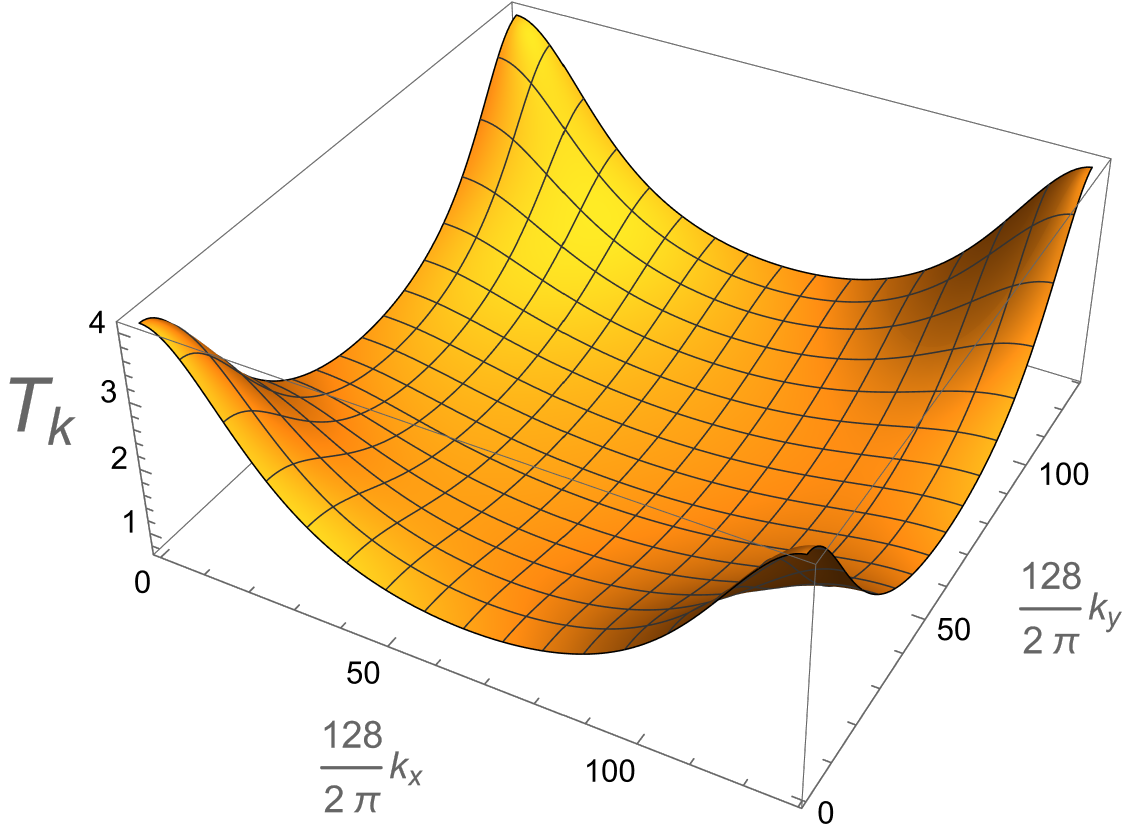}}\hspace{0.2cm}\subfloat[$\ln\left\langle \phi_{i}\phi_{0}\right\rangle $]{\centering{}\includegraphics[scale=0.33]{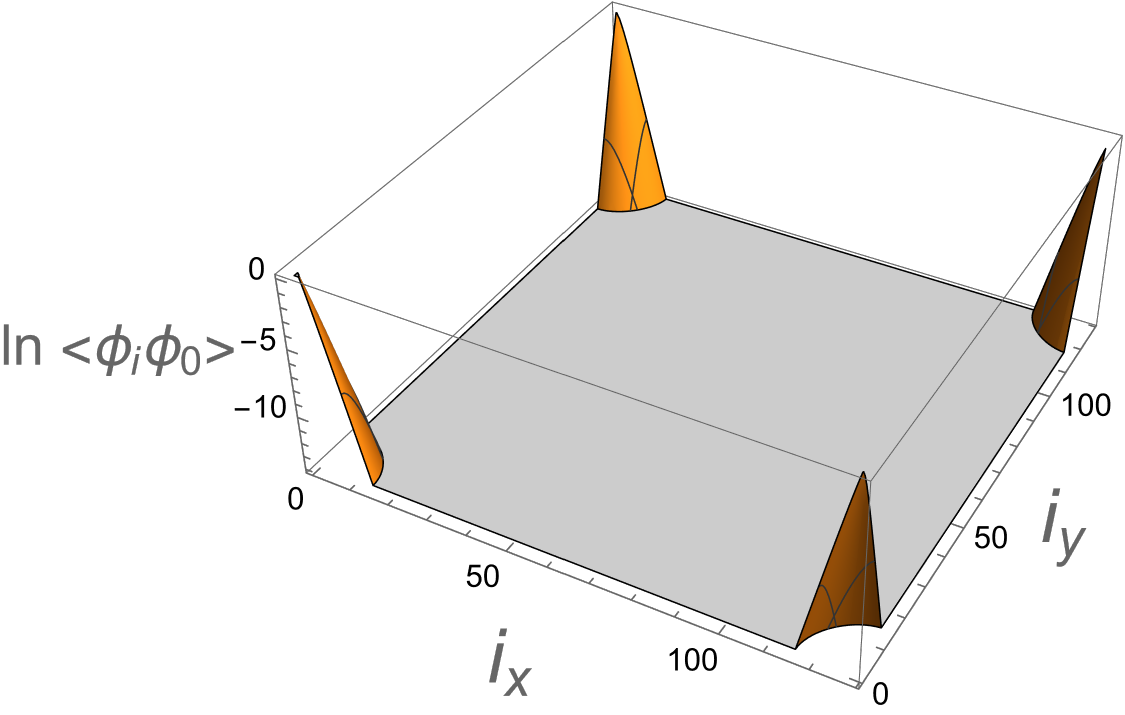}}\caption{Optimal $T_{k}$ and resulting real-space correlation $\left\langle \phi_{i}\phi_{0}\right\rangle $
(on log scale) for exemplary parameters $\left(a,b,\epsilon\right)=\left(-3,0.5,0.805\right)$.\label{fig:Optimal--and-1}}
\end{figure*}

A finite $128\times128$ lattice with periodic boundary conditions
was employed to declare the $T_{k}$ field. Partial derivatives of
$\mathcal{F}$ with respect to $T_{k}$ were calculated and used in
the Adam optimization algorithm \citep{Kingma}. It turns out, that
the algorithm does not converge and $\mathcal{F}$ is unbounded. The
reason is subtle and comes from the fact, that $\left\langle u^{2}\right\rangle $
was incorporated into $T_{k}$. $\left\langle u^{2}\right\rangle $
should be determined from the mean-field self-consistency condition,
which in this setting corresponds not to a minimum, but a maximum
of $\mathcal{F}\left(\left\langle u^{2}\right\rangle \right)$. After
fixing $\left\langle u^{2}\right\rangle $ according to this rule,
$\mathcal{F}$ is minimized for some $T_{k}$ (which now follows from
the variational principle). Therefore, in every step of the Adam algorithm,
$\left\langle u^{2}\right\rangle $ had to be adjusted to its self-consistent
value by solving $\partial\mathcal{F}\left(\left\langle u^{2}\right\rangle \right)/\partial\left\langle u^{2}\right\rangle =0$.
Of course instead of manipulating $\left\langle u^{2}\right\rangle $,
it can be fixed to $1$ and the global factor for $T_{k}$ can be
manipulated instead. These adjustments provided a well-behaved convergence.
Figures \ref{fig:Optimal--and} and \ref{fig:Optimal--and-1} show
the optimal $T_{k}$ field and the following real-space correlation
$\left\langle \phi_{i}\phi_{0}\right\rangle $ (under a logarithm)
for two different values of $\epsilon$. Low $\epsilon$ results in
a more disordered state. It can be seen that, as expected, $\left\langle \phi_{i}\phi_{0}\right\rangle $
decays linearly on a log scale with $\left|i\right|$ and this decay
is slower for higher $\epsilon$. $T_{k}$ becomes more concentrated
near low $k$ values for increased $\epsilon$.

\begin{figure}
\centering{}\includegraphics[scale=0.36]{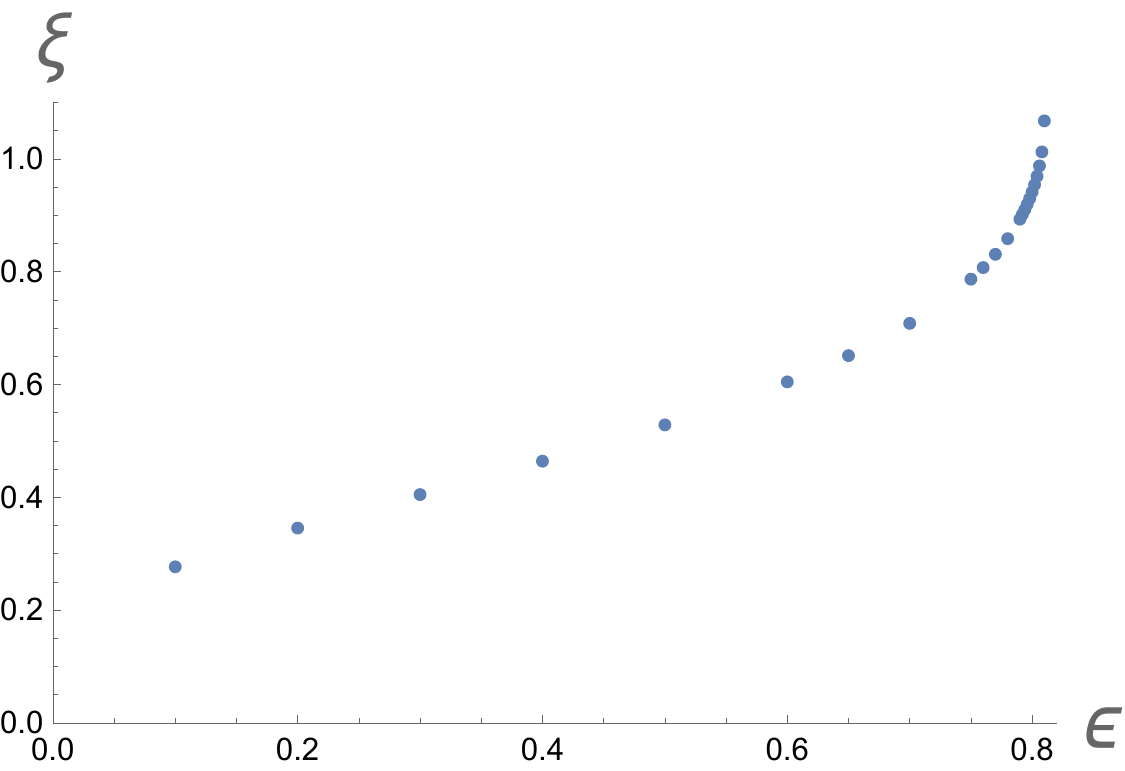}\caption{Correlation length as a function of $\epsilon$ for $\left(a,b\right)=\left(-3,0.5\right)$.\label{fig:Correlation-length-as}}
\end{figure}

Fitting a linear function to $\ln\left\langle \phi_{i}\phi_{0}\right\rangle $
allows to determine the correlation length $\xi$, plotted in Fig.
\ref{fig:Correlation-length-as} for various $\epsilon$. A sudden
increase in its value before $\epsilon_{c}\approx0.81$ indicates
the critical point. A natural question arises, whether critical exponents
can be determined in this approach. Numerical data from Fig. \ref{fig:Correlation-length-as}
is insufficient to find the critical exponent $\nu$, because maximal
$\xi$ is on the order of $1$. Achieving higher values is problematic,
because the algorithm reveals slower convergence in the vicinity of
the critical point. However, having an analytical expression for the
free energy, critical exponents can be attacked also analytically.

Equation \eqref{F/N(T)} can be rewritten as (with $f=\mathcal{F}/N$):

\begin{align}
f & =-\ln g\left(\frac{6b\left(x_{1}^{2}-x_{4}\right)-\epsilon x_{2}+ax_{1}}{\sqrt{bx_{4}}}\right)\nonumber \\
 & +\frac{1}{4}\ln x_{4}+3b\left(x_{4}-x_{1}^{2}\right)-x_{3}+\frac{1}{4}\ln b,
\end{align}
where:

\begin{equation}
g\left(x\right)=\int_{-\infty}^{\infty}\mathrm{d}u\,e^{-\left(u^{4}+xu^{2}\right)}.
\end{equation}

Equating a partial derivative of $f$ with respect to $T_{k}$ for
every $k$ to $0$ corresponds to finding an extremum of $f$. Since
$f$ depends on $T_{k}$ only through $x_{1},\dots,x_{4}$, it can
be expanded as:

\begin{equation}
\sum_{m}\frac{\partial f}{\partial x_{m}}\frac{\partial x_{m}}{\partial T_{k}}=0.\label{df/dTk}
\end{equation}
Additionally, if a global factor of $T_{k}$ is changed according
to the substitution $T_{k}\rightarrow cT_{k}$, then the derivative
of $f$ with respect to $c$ at $c=1$ also vanishes. This leads to
a condition:

\begin{equation}
-\left(\ln g\right)^{\prime}\left(\frac{6b\left(x_{1}^{2}-x_{4}\right)-\epsilon x_{2}+ax_{1}}{\sqrt{bx_{4}}}\right)=\sqrt{bx_{4}},
\end{equation}
which can be used to simplify \eqref{df/dTk} to the following form:

\begin{align}
 & \left(6bx_{1}+a-\epsilon J_{k}\right)T_{k}-\frac{1}{2}\frac{1}{T_{k}}\nonumber \\
 & -\frac{6bx_{1}^{2}-\epsilon x_{2}+ax_{1}-\frac{1}{2}}{x_{4}}\sum_{i}T_{i,0}^{3}e^{\mathrm{i}k\cdot i}=0.\label{eq for Tk}
\end{align}

Assuming standard asymptotic formulas for the correlation function
\citep{Binney} and the relation between $\left\langle \phi_{i}\phi_{j}\right\rangle $
and $T_{k}$ (Eq. \eqref{corr and Tk}) gives:

\begin{equation}
T_{k}\cong t^{-\gamma/2}\theta\left(\left|k\right|^{-1/\nu}t\right).\label{Tk}
\end{equation}
Equation \eqref{Tk} is valid for low $t$ (denoting the relative
distance to the critical point, for example $\left(\epsilon-\epsilon_{c}\right)/\epsilon_{c}$)
and low $\left|k\right|$. $\gamma$ and $\nu$ are the susceptibility-related
and correlation-length–related critical exponents. $\theta$ stands
for some single-argument function obeying the following asymptotic
relations:

\begin{equation}
\theta\left(x\right)\sim\begin{cases}
x^{\gamma/2} & \text{for }x\ll1\\
1 & \text{for }x\gg1
\end{cases}.
\end{equation}

Writing Eq. \eqref{eq for Tk} in an equivalent form:

\begin{equation}
T_{k}=\frac{\frac{6bx_{1}^{2}-\epsilon x_{2}+ax_{1}-\frac{1}{2}}{x_{4}}\sum_{i}T_{i,0}^{3}e^{\mathrm{i}k\cdot i}+\frac{1}{2}\frac{1}{T_{k}}}{6bx_{1}+a-\epsilon J_{k}},\label{final eq for Tk}
\end{equation}
asymptotic solution (Eq. \eqref{Tk}) can be substituted into it.
Numerical simulations suggest clearly that $x_{1}$, $\sum_{i}T_{i,0}^{3}$
and $x_{4}$ converge at the critical point. On the other hand, Eq.
\eqref{Tk} implies divergence of $T_{k=0}$. Examining behavior of
Eq. \eqref{final eq for Tk} near the critical point, it can be seen
that divergence of the denominator drives the phase transition. Exactly
at the critical point $6bx_{1}+a-\epsilon J_{k}$ scales as $\left|k\right|^{2}$
near $k=0$. If $\left(6bx_{1}^{2}-\epsilon x_{2}+ax_{1}-\frac{1}{2}\right)/x_{4}$
tends to a nonzero value, then $T_{k}\sim\left|k\right|^{-2}$. Then,
however, $x_{1}$ would not become a convergent integral in the thermodynamic
limit (in two dimensions). Even if $\left(6bx_{1}^{2}-\epsilon x_{2}+ax_{1}-\frac{1}{2}\right)/x_{4}$
tended to $0$, then $T_{k}\sim\left|k\right|^{-1}$, which generates
the same problem. The only way out of this apparent contradiction
is that $T_{k}$ does not have a well-defined limit at the critical
point. Therefore, scaling given by Eq. \eqref{Tk} seems not to be
properly reproduced by this method. While correlations, as opposed
to the mean-field approach, are accounted for and a sudden growth
of the correlation length $\xi$ is captured, the same cannot be said
about the nontrivial behavior of the system exactly at the critical
point. Critical exponent $\eta$ cannot be thus meaningfully obtained.
Regarding $\nu$, a nonobvious value may follow from Eq. \eqref{final eq for Tk},
but working it out is rather not easy (nor very useful).

\section{Percolation-based drawing for the Ising model\label{sec:Percolation-based-drawing-for}}

Both the $k$-space mean-field and $T$ matrix based method can be
viewed as independent drawing for each site followed by some linear
transformation of the $\phi$ vector. A natural question arises whether
every suitable (for the variational method) drawing has this structure.
To find the answer, a qualitatively different scheme should be analyzed.

For this purpose, similarity between the percolation and the Ising
model critical behavior can be used to solve (in the variational spirit)
the latter (assuming knowledge of the former). First, each bond is
set to be open independently with probability $p$. Open bonds form
clusters of sites. Then, within each cluster, all spins are set the
same, with equal probability of being $\pm1$. Such drawing is certainly
capable of reconstructing the nontrivial self-similar character of
the system near its critical point. This procedure is a sequence of
independent parts, so calculating the entropy should be easy. Yet,
the spins are correlated in a nonobvious way. The main problem with
it is that the same spin configuration can be achieved in different
ways. Therefore, their probabilities $p_{s}$ are not really simple
products (rather sums of products). Indeed, what is really drawn is
not the spin configuration itself, but a bonds plus spins configuration.
This is a major reason why a vast class of drawing schemes is actually
unsuitable for the variational method. However, it can be easily circumvented.

Suppose that percolation bonds are a part of the system. Since we
want to approximate the Gibbs distribution of the spins alone, bonds
can be governed by any convenient distribution. In other words, the
probability $P_{1}\left(p;b,s\right)$ of drawing bond configuration
$b$ and spin configuration $s$ (using percolation parameter $p$)
should approximate $P_{\mathrm{Gibbs}}\left(s\right)P_{2}\left(q,s;b\right)$,
which is a probability of getting spin state $s$ from the Gibbs distribution
and then bond configuration $b$ drawn subsequently with probability
$P_{2}$ depending on $s$ and some variable parameters $q$. This
can be schematically written as:

\begin{equation}
P_{1}\left(p;b,s\right)\cong P_{\mathrm{Gibbs}}\left(s\right)P_{2}\left(q,s;b\right).\label{ult var meth}
\end{equation}

The Gibbs distribution is given by the Boltzmann factor normalized
appropriately by the true free energy $\mathcal{F}_{\mathrm{true}}$:

\begin{equation}
P_{\mathrm{Gibbs}}\left(s\right)=e^{\beta\left(\mathcal{F}_{\mathrm{true}}-\mathcal{H}\left(s\right)\right)}.\label{Gibbs}
\end{equation}
Substituting Eq. \eqref{Gibbs} into Eq. \eqref{ult var meth} and
bringing $P_{2}$ into the exponent gives:

\begin{equation}
P_{1}\left(p;b,s\right)\cong e^{\beta\left(\mathcal{F}_{\mathrm{true}}-\tilde{\mathcal{H}}\left(b,s\right)\right)},\label{approx}
\end{equation}
where:

\begin{equation}
\tilde{\mathcal{H}}\left(b,s\right)=\mathcal{H}\left(s\right)-\frac{1}{\beta}\ln P_{2}\left(q,s;b\right).\label{H tilde}
\end{equation}
Therefore, $P_{1}$ is meant to approximate a distribution dictated
by Hamiltonian $\tilde{\mathcal{H}}$, which generates free energy
$\mathcal{F}_{\mathrm{true}}$ (indeed, the right-hand-side of Eq.
\eqref{approx} represents a normalized distribution, since the right-hand-side
of Eq. \eqref{ult var meth} is normalized). Variational free energy
$\mathcal{F}\left(p,q\right)$ calculated for $P_{1}$ and $\tilde{\mathcal{H}}$
is given by:

\begin{align}
\mathcal{F}\left(p,q\right) & =\sum_{bs}P_{1}\left(p;b,s\right)\tilde{\mathcal{H}}\left(b,s\right)\nonumber \\
 & +\frac{1}{\beta}\sum_{bs}P_{1}\left(p;b,s\right)\ln P_{1}\left(p;b,s\right).
\end{align}
Inserting Eq. \eqref{H tilde} leads to:

\begin{align}
\mathcal{F}\left(p,q\right) & =\sum_{bs}P_{1}\left(p;b,s\right)\mathcal{H}\left(s\right)\nonumber \\
 & -\frac{1}{\beta}\sum_{bs}P_{1}\left(p;b,s\right)\ln P_{2}\left(q,s;b\right)\nonumber \\
 & +\frac{1}{\beta}\sum_{bs}P_{1}\left(p;b,s\right)\ln P_{1}\left(p;b,s\right).\label{F}
\end{align}

According to the variational principle \citep{Falk}, $\mathcal{F}\left(p,q\right)\geq\mathcal{F}_{\mathrm{true}}$.
Thus, parameters $p$ and $q$ should be optimized to minimize $\mathcal{F}\left(p,q\right)$.
Term in the second line in Eq. \eqref{F} (multiplied by $\beta$)
can be named ``correction entropy'', because without it the expression
resembles standard variational formula. The latter however, involves
probabilities of single bond-spin configurations instead of spin configurations,
which significantly overestimates the entropy. The correction entropy
accounts for this fact having opposite sign to the last term in Eq.
\eqref{F}.

Probability distribution $P_{2}$ has to be designed in such a way
that the correction entropy can be analytically handled. Ideally,
it would equal the conditional probability of getting bond configuration
$b$ given spin configuration $s$. However, this is not easy to have
these two features simultaneously. At least, a sensible $P_{2}$ distribution
should generate only such $b$, which are compatible with $s$ (i.
e. $P_{1}\left(p;b,s\right)\neq0$). A possible design is as follows.
For given $s$ each bond is set open independently, with probability
$q$ if it links aligned spins and probability $0$ if it links opposite
spins.

Now the variational free energy can be determined for the Ising model
Hamiltonian:

\begin{equation}
\mathcal{H}=-\epsilon\sum_{\left\langle ij\right\rangle }s_{i}s_{j}\label{Ising H}
\end{equation}
and described drawing schemes.

The mean energy $\sum_{bs}P_{1}\left(p;b,s\right)\mathcal{H}\left(s\right)$
equals $-\epsilon Nz\left\langle s_{i}s_{j}\right\rangle $ (for neighboring
$i,j$). Let $P\left(p\right)$ be the probability that neighboring
sites are joined by a path in the percolation model with occupation
probability $p$. Then $\left\langle s_{i}s_{j}\right\rangle =P\left(p\right)$.
The entropy term (the last in Eq. \eqref{F}) can be calculated as
($-\beta^{-1}$ times) a sum of entropies associated with subsequent
steps in the drawing scheme. Generating the bond configuration brings
entropy

\begin{equation}
-\frac{Nz}{2}\left[p\ln p+\left(1-p\right)\ln\left(1-p\right)\right],\label{S}
\end{equation}
while generating spins for given $b$ produces entropy $NM\left(p\right)\ln2$.
$M\left(p\right)$ is the mean number of clusters per $N$ in the
percolation model. The correction entropy $S_{\mathrm{corr}}$ is
an averaged value of $-\ln P_{2}$ with respect to distribution $P_{1}$,
so:

\begin{align}
 & S_{\mathrm{corr}}=-\left\langle \#\text{links connecting aligned spins}\right\rangle \ln q\nonumber \\
 & -\left\langle \#\text{missing links between aligned spins}\right\rangle \ln\left(1-q\right).\label{Scorr}
\end{align}

Spins connected by a percolation link are always aligned, so the first
average in Eq. \eqref{Scorr} is simply the total number of links,
which equals $Npz/2$. The second average equals $Nz/2$ times the
probability that a bond is not open (a missing link) and spins at
its ends are additionally aligned. Let $P_{\mathrm{indir}}\left(p\right)$
be a probability that if neighboring sites are not joined directly,
they are join indirectly by a percolation path. Then:

\begin{equation}
P\left(p\right)=p+\left(1-p\right)P_{\mathrm{indir}}\left(p\right)\label{Pp}
\end{equation}
and

\begin{align}
 & \left\langle \#\text{missing links between aligned spins}\right\rangle =\nonumber \\
 & =\frac{Nz}{2}\left(1-p\right)\left[P_{\mathrm{indir}}\left(p\right)+\frac{1}{2}\left(1-P_{\mathrm{indir}}\left(p\right)\right)\right].\label{links}
\end{align}

Using Eq. \eqref{Pp} to write $P_{\mathrm{indir}}\left(p\right)$
in terms of $P\left(p\right)$ and substituting all intermediate results
into the structure of Eq. \eqref{F} gives the variational free energy
per site:

\begin{align}
 & \frac{\mathcal{F}\left(p,q\right)}{N}=-\epsilon zP\left(p\right)\nonumber \\
 & -\frac{z}{2\beta}\left[p\ln q+\frac{1}{2}\left(1+P\left(p\right)-2p\right)\ln\left(1-q\right)\right]\nonumber \\
 & +\frac{z}{2\beta}\left[p\ln p+\left(1-p\right)\ln\left(1-p\right)\right]-\frac{1}{\beta}M\left(p\right)\ln2.\label{f}
\end{align}

Optimization over $q$ can be performed exactly, because it enters
only the correction entropy in a simple way. Writing the necessary
condition for a minimum:

\begin{equation}
\frac{\mathrm{d}}{\mathrm{d}q}\left[C_{1}\ln q+C_{2}\ln\left(1-q\right)\right]=\frac{C_{1}}{q}-\frac{C_{2}}{1-q}=0,
\end{equation}
leads to an optimal value of $q$:

\begin{equation}
q=\frac{2p}{1+P\left(p\right)}.
\end{equation}
Substituting it to Eq. \eqref{f} and simplifying gives:

{\small
\begin{align}
 & \frac{\mathcal{F}\left(p\right)}{N}=-\epsilon zP\left(p\right)-\frac{1}{\beta}M\left(p\right)\ln2\nonumber \\
 & +\frac{z}{2\beta}\left[\left(1-p\right)\ln\left(1-p\right)-\frac{1+P\left(p\right)-2p}{2}\ln\frac{1+P\left(p\right)-2p}{2}\right.\nonumber \\
 & \left.+\frac{1+P\left(p\right)}{2}\ln\frac{1+P\left(p\right)}{2}\right].\label{f-1}
\end{align}
}{\small\par}

\begin{figure*}
\centering{}\subfloat[Two dimensional case]{\centering{}\includegraphics[scale=0.31]{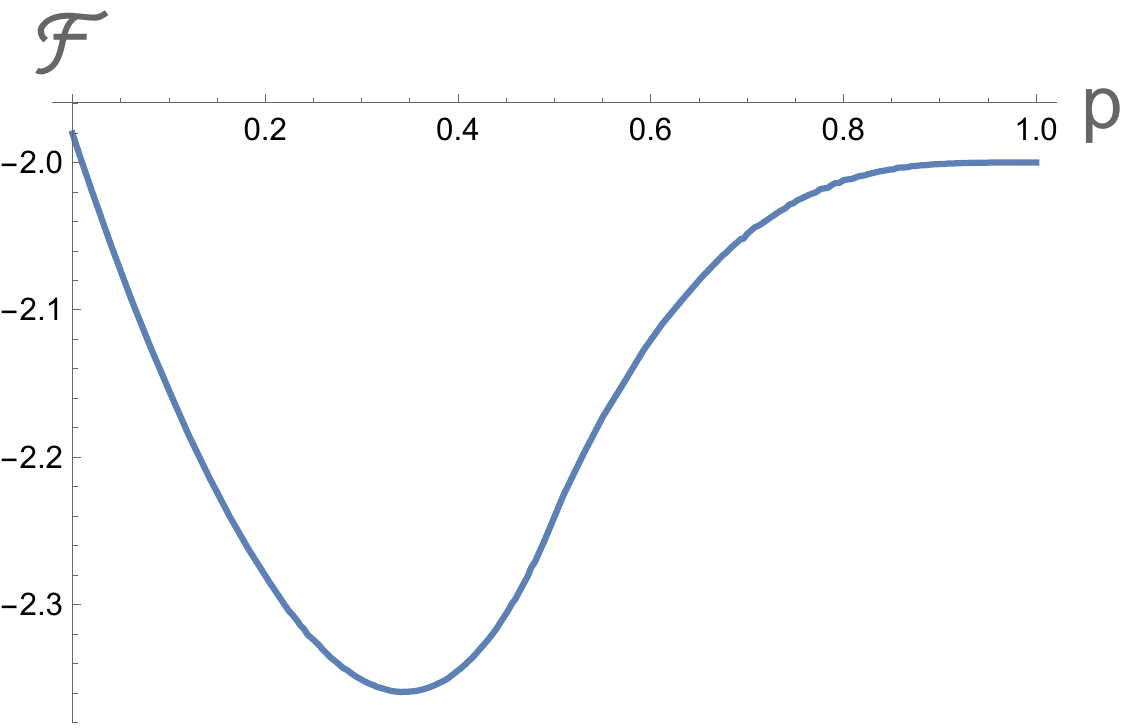}}\hspace{0.2cm}\subfloat[Three dimensional case]{\centering{}\includegraphics[scale=0.31]{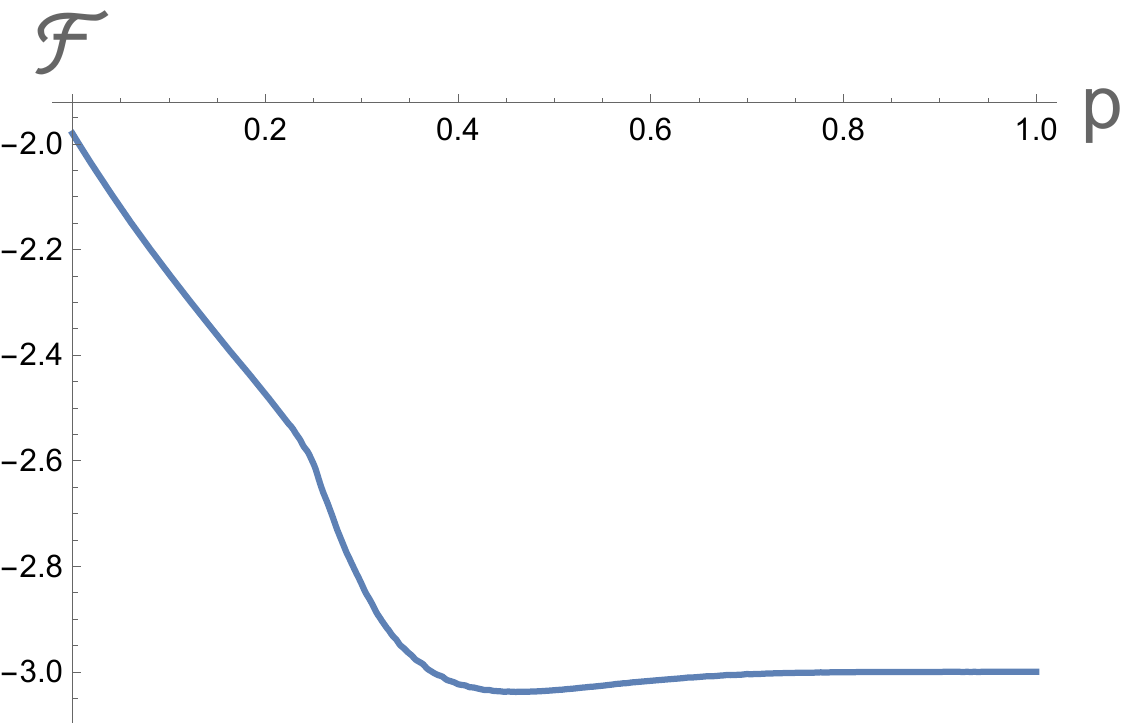}}\caption{Variational free energy as a function of the percolation parameter
$p$ for an exemplary inverse temperature $\beta=0.35$.\label{fig:Variational-free-energy}}
\end{figure*}

\begin{figure*}
\centering{}\subfloat[Two dimensional case]{\centering{}\includegraphics[scale=0.34]{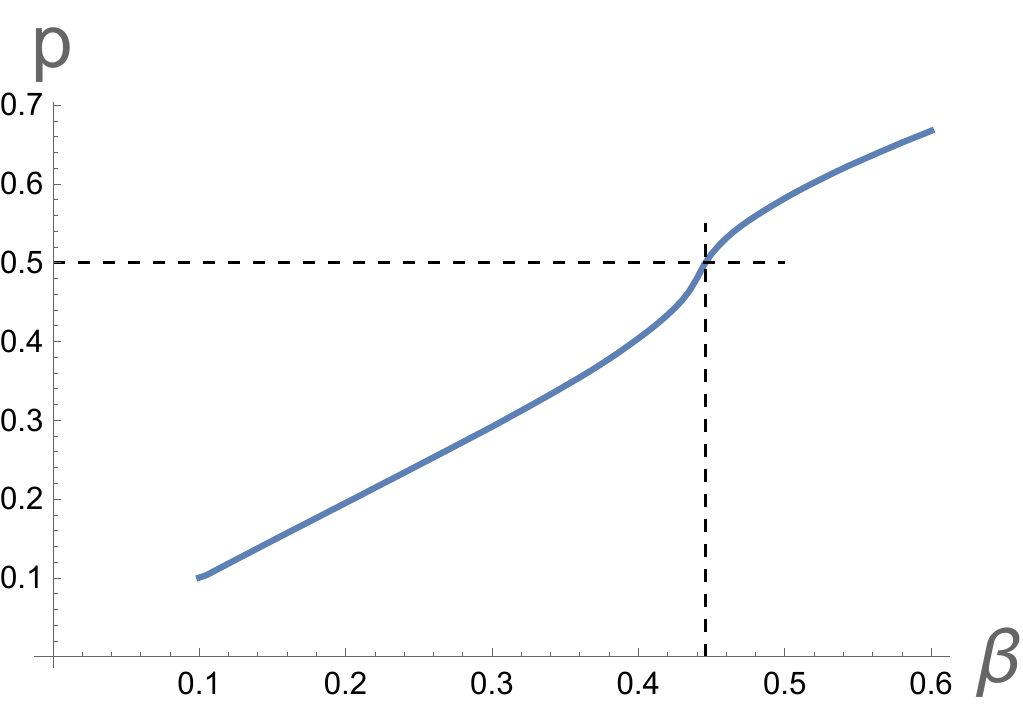}}\hspace{0.2cm}\subfloat[Three dimensional case]{\centering{}\includegraphics[scale=0.34]{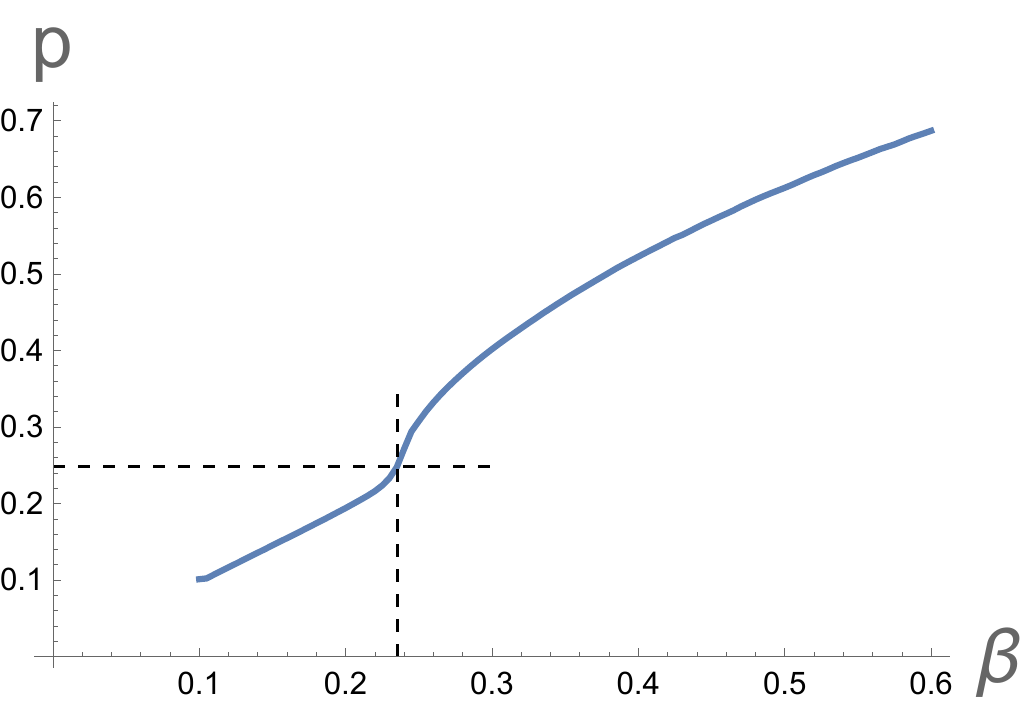}}\caption{Optimal variational parameter $p\left(\beta\right)$ as a function
of the inverse temperature with marked corresponding critical values.\label{fig:Optimal-variational-parameter}}
\end{figure*}

Knowledge of functions $P\left(p\right)$ and $M\left(p\right)$ is
assumed. For the purposes of this work they are determined using the
Newman-Ziff algorithm \citep{Newman} written in Python. The two dimensional
system was simulated on a lattice $500\times500$ with $50$ repetitions,
while the three dimensional on $100\times100\times100$ also with
$50$ repetitions. Figure \ref{fig:Variational-free-energy} shows
exemplary plots of $\mathcal{F}\left(p\right)$ (normalized by $N$)
in two and three dimensions. $\epsilon$ is set to $0.5$, because
it corresponds to having $\epsilon=1$ and summing only over unordered
pairs in Eq. \eqref{Ising H}, which is the convention taken by Binney
\citep{Binney} (this reference is used for comparison). For every
$\beta$, optimal value of the percolation parameter can be found,
which results in a function $p\left(\beta\right)$. Critical value
$p_{c}$ known from percolation satisfies $p_{c}=p\left(\beta_{c}\right)$
for the sought critical inverse temperature of the Ising model. This
can be justified in a number of ways, for example by noting that the
mean magnetization $\left\langle s\right\rangle $ (assuming the spin-flip
$\mathbb{Z}_{2}$ symmetry is spontaneously broken) equals the probability
$P_{\infty}$ that a site belongs to the infinite cluster. Both quantities
share qualitatively similar behavior \citep{Binney,Saberi} with a
sharp onset point (the critical point). Figure \ref{fig:Optimal-variational-parameter}
presents $p\left(\beta\right)$ plots for $d=2,3$. In these dimensionalities
the percolation thresholds are $p_{c}=1/2$ and $p_{c}=0.2488$ respectively
\citep{Stover}. Corresponding critical inverse temperatures for the
Ising model are $\beta_{c}=0.446$ ($d=2$) and $\beta_{c}=0.235$
($d=3$). The literature values are $\beta_{c}=0.4407$ (from exact
solution in $d=2$) and $\beta_{c}=0.222$ ($d=3$) \citep{Binney}.
This means discrepancy of $1.2\,\%$ and $5.9\,\%$ respectively.

One may ask about the physical interpretation of Eq. \eqref{ult var meth}.
Although it can be regarded merely as a mathematical trick, in which
the configuration space is artificially extended to allow easier fitting,
it is worth providing a physical view. If two physical entities lack
direct interaction, they still can interact indirectly via mediation
of other particles. This is very cleanly exemplified by the RKKY interaction
\citep{Kittel}. Magnetic moments become coupled to one another by
conduction electrons. If one forgets about the electrons (for example
integrating them out in the path integral formalism), it seems as
there was an interaction between the magnetic moments. Equation \eqref{ult var meth}
does a kind of reverse operation. The variational drawing scheme cannot
easily mimic direct interactions between the spins, so it produces
them by immersing non-interacting spins in a bath of bonds governed
by a pristine percolation model. The ``variational fitting'' (i.
e. minimization of the free energy) has to be performed in the extended
reality (i. e. with the bonds present), so that it can be technically
performed. Thus, the system of spins governed by the original Hamiltonian
has to be enriched by bonds via including $P_{2}$. Both $P_{1}$
and $P_{2}$ are optimized (by adjusting parameters $p$ and $q$),
so physically, the method finds the best possible bath of quasi-particles
(with simple behavior on their own, here realized by the bonds) that
the interaction between spins can be understood as an indirect one
mediated by the bath.

Further insight can be provided by comparing the method of this section
with the Fortuin-Kasteleyn representation of the Ising model \citep{Saberi,Kasteleyn,Fortuin}.
The latter consists in defining the ($\kappa=2$) random cluster model
\citep{Fortuin}, which assigns a given bond configuration (with $N_{B}$
bonds and $N_{C}$ clusters) probability proportional to
\begin{equation}
\left(1-p\right)^{Nz/2-N_{B}}p^{N_{B}}2^{N_{C}}.\label{factor}
\end{equation}
The factor of $2^{N_{C}}$ differentiates the model from ordinary
percolation. Then, setting spins to be identical within each cluster
and equally likely up or down, produces the true Ising model distribution
at inverse temperature $\beta$ satisfying $p=1-e^{-2\beta\epsilon}$.

In the presented variational method, the Ising model distribution
of spin configurations cannot be reconstructed exactly, because instead
of the random cluster model only standard percolation is used for
drawing bonds (formally a $\kappa=1$ random cluster model). However,
shifting $p$ away from $1-e^{-2\beta\epsilon}$ can mimic the effect
of the $2^{N_{C}}$ factor in formula \eqref{factor}. Variational
method gives the best possible choice as $p\left(\beta\right)$, plotted
in Fig. \ref{fig:Optimal-variational-parameter}.

The relation between standard percolation and the Ising model is not
fully straightforward, but can be established in an approximate manner
using the variational method. Different connections were found (among
many others) by Bishop \citep{Bishop}, who finds approximate relations
between percolation and the Ising model by examining Bethe lattices
or Hu \citep{Hu} finding an exact relation, but for a bond-correlated
percolation model (essentially the $\kappa=2$ random-cluster model).

\section{Fractal-like drawing reproducing RG-style results\label{sec:Fractal-like-drawing-reproducing}}

Partial failure of the method from section \ref{sec:Including-correlations-as},
regarding describing the critical behavior, can be understood from
the perspective of renormalization group methods \citep{Wilson1,Wilson2,Kadanoff}.
At the critical point a system exhibits not only long-range correlations
(which matrix $T$ could account for), but also self-similar fractal
character, which was absent in that drawing scheme. Using the general
method for handling multi-stage drawing procedures from the preceding
section, a fractal-like scheme can be now developed. This approach
is inspired by the renormalization group, to which it is closely related.
However, it is conceptually and practically different from the variational
renormalization group of Kadanoff \citep{Kadanoff}. A more detailed
comparison is given at the end of this section.

Again, the Ising model is used for illustrative purposes. We grow
a spin configuration iteratively starting from a single spin. Then,
a procedure inverse to renormalization is needed to generate more
and finally all $N$ sites. This process can be called ``up-normalization''
(because the number of sites is upscaled). It consist of a drawing
recipe how to replace each spin by a cluster of spins. Similarly to
the percolation-based drawing, not only a final configuration is generated
by also all intermediate states. Thus, a trick introduced in Eq. \eqref{ult var meth}
is needed. This requires a recipe for generating intermediate configurations
from the final one, which is exactly the standard renormalization
reduction in the number of degrees of freedom. In analogy to ``up-normalization''
this procedure can be coined ``down-normalization''.

Standardly \citep{Binney}, let $b^{d}$ (where $d$ is dimensionality)
denote the number of spins generated from one in up-normalization.
Then, down-normalization reduces a cluster of $b^{d}$ spins to a
single one. Let $s^{\left(m\right)}$ be a spin configuration obtained
in the $m$-th step and $s^{\left(n\right)}$ be the final configuration
meant to approximate the true Gibbs distribution. Let $\mathrm{UN}\left(p,s^{\left(m-1\right)};s^{\left(m\right)}\right)$
denote the probability of generating a spin configuration $s^{\left(m\right)}$,
given the preceding one $s^{\left(m-1\right)}$ and a set of adjustable
parameters $p$ present in the up-normalization recipe. Similarly,
down-normalization can be given by probability $\mathrm{DN}\left(q,s^{\left(m\right)};s^{\left(m-1\right)}\right)$
of down-normalizing to $s^{\left(m-1\right)}$ from $s^{\left(m\right)}$
with given adjustable parameters $q$. In each step we can use different
adjustable parameters, with one goal to approximate the Gibbs distribution
as accurately as possible at the final step. Therefore, the analog
of Eq. \eqref{ult var meth} becomes:

\begin{align}
 & \prod_{m=1}^{n}\mathrm{UN}\left(p^{\left(m\right)},s^{\left(m-1\right)};s^{\left(m\right)}\right)\nonumber \\
 & \cong P_{\mathrm{Gibbs}}\left(s^{\left(n\right)}\right)\prod_{m=1}^{n}\mathrm{DN}\left(q^{\left(m\right)},s^{\left(m\right)};s^{\left(m-1\right)}\right).\label{UN, DN}
\end{align}

Repeating the reasoning from the previous section, Eq. \eqref{UN, DN}
leads to the following expression for the free energy (analogous to
Eq. \eqref{F}):

\begin{align}
 & \mathcal{F}\left(p,q\right)\nonumber \\
 & =\left\langle \mathcal{H}\left(s^{\left(n\right)}\right)\right\rangle -\frac{1}{\beta}\sum_{m=1}^{n}\left\langle \ln\mathrm{DN}\left(q^{\left(m\right)},s^{\left(m\right)};s^{\left(m-1\right)}\right)\right\rangle \nonumber \\
 & +\frac{1}{\beta}\sum_{m=1}^{n}\left\langle \ln\mathrm{UN}\left(p^{\left(m\right)},s^{\left(m-1\right)};s^{\left(m\right)}\right)\right\rangle .\label{F-1}
\end{align}
Averaging is carried over the proposed drawing scheme (which is shaped
only by $\mathrm{UN}$). Again, mean energy, correction entropy and
entropy terms can be recognized in Eq. \eqref{F-1}.

To exemplify the given approach a very crude up and down-normalization
schemes are to be used in two dimensions. The first takes only one
parameter $p\in\left[0,\frac{1}{2}\right]$ and replaces a single
spin by a $2\times2$ block (so $b=2$). Each of the newly generated
spins have initially the orientation of its parent and then is independently
flipped with probability $p$. The down-normalization has to draw
a $\pm1$ value for every possible $2\times2$ block. Its algorithm
can be proposed to be as follows.

If all spins in a block are identically aligned, the same orientation
is finally chosen with probability $q_{0}$ (and the opposite with
probability $1-q_{0}$). If only a single spin is misaligned, the
majority orientation is taken with probability $q_{1}$ (and the opposite
with probability $1-q_{1}$). If two spin are $-1$ and the other
two are $+1$, orientation is drawn uniformly.

We take the Ising Hamiltonian from Eq. \eqref{Ising H} extended by
the magnetic field term, resulting in:

\begin{equation}
\mathcal{H}=-\epsilon\sum_{\left\langle ij\right\rangle }s_{i}s_{j}-B\sum_{i}s_{i}.\label{H-1}
\end{equation}

The mean energy from Eq. \eqref{F-1} is:

\begin{equation}
\left\langle \mathcal{H}\left(s^{\left(n\right)}\right)\right\rangle =-\epsilon Nz\left\langle s_{i}^{\left(n\right)}s_{j}^{\left(n\right)}\right\rangle -BN\left\langle s_{i}^{\left(n\right)}\right\rangle ,
\end{equation}
where $i,j$ is a pair of nearest neighbors. $\left\langle s_{i}^{\left(m\right)}s_{j}^{\left(m\right)}\right\rangle $
and $\left\langle s_{i}^{\left(m\right)}\right\rangle $ can be determined
recursively on the basis of the up-normalization scheme. In every
iteration a spin $+1$ is replaced by four independently drawn spins,
each with average value $\left(1-p\right)-p=1-2p$. Thus:

\begin{equation}
\left\langle s_{i^{\prime}}^{\left(m+1\right)}\right\rangle =\left(1-2p^{\left(m+1\right)}\right)\left\langle s_{i}^{\left(m\right)}\right\rangle .\label{rec1}
\end{equation}
$i^{\prime}$ is just a site from the $m+1$-th generation (as opposed
to $i$ belonging to the $m$-th generation). If a bond is chosen
at random in the $m+1$-th generation, there is $1/2$ chance that
it connects spins from the same block (i. e. generated from a single
spin from the $m$-th generation). If this possibility occurs, then:

\begin{align}
 & \left\langle s_{i^{\prime}}^{\left(m+1\right)}s_{j^{\prime}}^{\left(m+1\right)}\right\rangle _{\text{case1}}\nonumber \\
 & =p^{\left(m+1\right)2}+\left(1-p^{\left(m+1\right)}\right)^{2}-2p^{\left(m+1\right)}\left(1-p^{\left(m+1\right)}\right)\nonumber \\
 & =\left(1-2p^{\left(m+1\right)}\right)^{2}.
\end{align}

In the remaining case, the calculation is analogous, but an additional
factor of $\left\langle s_{i}^{\left(m\right)}s_{j}^{\left(m\right)}\right\rangle $
(where sites $i^{\prime}$ and $j^{\prime}$ generated $i$ and $j$
in up-normalization) appears due to averaging over possible orientations
of the parent spins:

\begin{equation}
\left\langle s_{i^{\prime}}^{\left(m+1\right)}s_{j^{\prime}}^{\left(m+1\right)}\right\rangle _{\text{case2}}=\left(1-2p^{\left(m+1\right)}\right)^{2}\left\langle s_{i}^{\left(m\right)}s_{j}^{\left(m\right)}\right\rangle .
\end{equation}

Finally, averaging over the two cases gives:

\begin{equation}
\left\langle s_{i^{\prime}}^{\left(m+1\right)}s_{j^{\prime}}^{\left(m+1\right)}\right\rangle =\frac{\left(1-2p^{\left(m+1\right)}\right)^{2}}{2}\left(1+\left\langle s_{i}^{\left(m\right)}s_{j}^{\left(m\right)}\right\rangle \right).\label{rec2}
\end{equation}

The entropy term is straightforward:

\begin{align}
 & \frac{1}{\beta}\sum_{m=1}^{n}\left\langle \ln\mathrm{UN}\left(p^{\left(m\right)},s^{\left(m-1\right)};s^{\left(m\right)}\right)\right\rangle =\nonumber \\
 & =\frac{1}{\beta}\sum_{m=1}^{n}N_{m}\left[p^{\left(m\right)}\ln p^{\left(m\right)}+\left(1-p^{\left(m\right)}\right)\ln\left(1-p^{\left(m\right)}\right)\right],\label{entropy term}
\end{align}
where $N_{m}=N/2^{\left(n-m\right)d}$ denotes the number of sites
in the $m$-th stage. Similarly, the correction entropy term becomes:

\begin{align}
 & -\frac{1}{\beta}\sum_{m=1}^{n}\left\langle \ln\mathrm{DN}\left(q^{\left(m\right)},s^{\left(m\right)};s^{\left(m-1\right)}\right)\right\rangle \nonumber \\
 & =-\frac{1}{\beta}\sum_{m=1}^{n}N_{m-1}\left[\left(1-p^{\left(m\right)}\right)^{4}\ln\left(1-q_{0}^{\left(m\right)}\right)\right.\nonumber \\
 & +4p^{\left(m\right)}\left(1-p^{\left(m\right)}\right)^{3}\ln\left(1-q_{1}^{\left(m\right)}\right)\nonumber \\
 & +6p^{\left(m\right)2}\left(1-p^{\left(m\right)}\right)^{2}\ln\frac{1}{2}\nonumber \\
 & \left.+4p^{\left(m\right)3}\left(1-p^{\left(m\right)}\right)\ln q_{1}^{\left(m\right)}+p^{\left(m\right)4}\ln q_{0}^{\left(m\right)}\right].\label{corr entropy}
\end{align}

Coefficients $\left(1-p^{\left(m\right)}\right)^{4}$, $4p^{\left(m\right)}\left(1-p^{\left(m\right)}\right)^{3}$,
$6p^{\left(m\right)2}\left(1-p^{\left(m\right)}\right)^{2}$, $4p^{\left(m\right)3}\left(1-p^{\left(m\right)}\right)$
and $p^{\left(m\right)4}$ are probabilities that a given parent spin
transforms into exactly $4$, $3$, $2$, $1$ and $0$ spins aligned
like it respectively. Optimization over $q$ can be performed immediately
on the basis of the following formula:

\begin{align}
 & \max_{x}\left\{ A_{1}\ln x+A_{2}\ln\left(1-x\right)\right\} \nonumber \\
 & =A_{1}\ln A_{1}+A_{2}\ln A_{2}-\left(A_{1}+A_{2}\right)\ln\left(A_{1}+A_{2}\right),
\end{align}
 for $A_{1},A_{2}\in\mathbb{R}_{+}$. Thus the correction entropy
and entropy terms together can be written as

\begin{equation}
\frac{N}{\beta}\sum_{m=1}^{n}\frac{g\left(p^{\left(m\right)}\right)}{4^{\left(n-m\right)}},
\end{equation}
where:

\begin{align}
g\left(x\right) & =\phi\left(x\right)+\phi\left(1-x\right)\nonumber \\
 & -\left(1-x\right)^{3}\phi\left(1-x\right)-x^{3}\phi\left(x\right)\nonumber \\
 & +\frac{1}{4}\phi\left(\left(1-x\right)^{4}+x^{4}\right)\nonumber \\
 & +\frac{3}{2}x^{2}\left(1-x\right)^{2}\ln2-2x\left(1-x\right)\nonumber \\
 & \times\left[\left(1-x\right)\phi\left(1-x\right)+x\phi\left(x\right)\right.\nonumber \\
 & \left.-\frac{1}{2}\phi\left(\left(1-x\right)^{2}+x^{2}\right)\right],
\end{align}
\begin{equation}
\phi\left(x\right)=x\ln x.
\end{equation}
Expression for $g\left(x\right)$ is lengthy, but extremely simple
from a computational point of view and fully analytical. Finally,
the free energy per site takes form:

\begin{align}
f\left(p\right) & =-\epsilon z\left\langle s_{i}^{\left(n\right)}s_{j}^{\left(n\right)}\right\rangle -B\left\langle s_{i}^{\left(n\right)}\right\rangle \nonumber \\
 & +\frac{1}{\beta}\sum_{m=1}^{n}\frac{g\left(p^{\left(m\right)}\right)}{4^{\left(n-m\right)}}.\label{f-2}
\end{align}

Parameters $p^{\left(m\right)}$ minimizing $f\left(p\right)$ provide
the best approximation to the Gibbs distribution for the proposed
drawing scheme. Let us split the optimization problem into two parts,
first over $p^{\left(1\right)},\cdots,p^{\left(n-1\right)}$ and finally
over $p^{\left(n\right)}$. Using the recursion rules from Eqs. \eqref{rec1}
and \eqref{rec2} in Eq. \eqref{f-2} gives:

\begin{align}
f\left(p\right) & =-\bar{\epsilon}z\left\langle s_{i}^{\left(n-1\right)}s_{j}^{\left(n-1\right)}\right\rangle -\bar{B}\left\langle s_{i}^{\left(n-1\right)}\right\rangle \nonumber \\
 & +\frac{1}{\bar{\beta}}\sum_{m=1}^{n-1}\frac{g\left(p^{\left(m\right)}\right)}{4^{\left(n-1-m\right)}}+\frac{1}{\beta}g\left(p^{\left(n\right)}\right)-\epsilon z\frac{\left(1-2p^{\left(n\right)}\right)^{2}}{2},\label{f-2-1}
\end{align}
where:

\begin{equation}
\begin{cases}
\bar{\epsilon}=\frac{\left(1-2p^{\left(n\right)}\right)^{2}}{2}\epsilon\\
\bar{B}=\left(1-2p^{\left(n\right)}\right)B\\
\bar{\beta}=4\beta
\end{cases}.\label{bar}
\end{equation}

The task of minimizing over $p^{\left(1\right)},\cdots,p^{\left(n-1\right)}$
is now equivalent to solving the full optimization problem, but with
changed values of $\epsilon,B,\beta$ and $n$. The last quantity
is changed to $n-1$. Writing truly all arguments of $f\left(p\right)$
in Eq. \eqref{f-2-1} it becomes $f_{n}\left(\epsilon,B,\beta,p^{\left(1\right)},\cdots,p^{\left(n\right)}\right)$.
Then:

\begin{align}
 & f_{n}\left(\epsilon,B,\beta,p^{\left(1\right)},\cdots,p^{\left(n\right)}\right)\nonumber \\
 & =f_{n-1}\left(\bar{\epsilon},\bar{B},\bar{\beta},p^{\left(1\right)},\cdots,p^{\left(n-1\right)}\right)\nonumber \\
 & +\frac{1}{\beta}g\left(p^{\left(n\right)}\right)-\epsilon z\frac{\left(1-2p^{\left(n\right)}\right)^{2}}{2}.
\end{align}

Scaling in temperature can be eliminated by using a generally valid
scaling

\begin{align}
 & f\left(c^{-1}\epsilon,c^{-1}B,c\beta\right)\nonumber \\
 & =-\frac{1}{Nc\beta}\ln\mathrm{Tr}e^{-\beta\mathcal{H}}=\frac{1}{c}f\left(\epsilon,B,\beta\right),
\end{align}
which leads to:

\begin{align}
 & f_{n}\left(\epsilon,B,\beta,p^{\left(1\right)},\cdots,p^{\left(n\right)}\right)\nonumber \\
 & =\frac{1}{4}f_{n-1}\left(\epsilon^{\prime},B^{\prime},\beta,p^{\left(1\right)},\cdots,p^{\left(n-1\right)}\right)\nonumber \\
 & +\frac{1}{\beta}g\left(p^{\left(n\right)}\right)-\epsilon z\frac{\left(1-2p^{\left(n\right)}\right)^{2}}{2},\label{f recur}
\end{align}
with $\epsilon^{\prime}=4\bar{\epsilon}$ and $B^{\prime}=4\bar{B}$.
Then using Eq. \eqref{bar} one receives:

\begin{equation}
\begin{cases}
\epsilon^{\prime}=2\left(1-2p^{\left(n\right)}\right)^{2}\epsilon\\
B^{\prime}=4\left(1-2p^{\left(n\right)}\right)B
\end{cases}.\label{final rec}
\end{equation}

Equation \eqref{f recur} has a natural interpretation in the spirit
of the renormalization group. Since only $f_{n-1}$ contains parameters
$p^{\left(1\right)},\cdots,p^{\left(n-1\right)}$ on the right-hand-side,
their optimal values provide possibly the best Gibbs distribution
for a Hamiltonian $\mathcal{H}^{\prime}$ (defined by Eq. \eqref{H-1},
but with $\epsilon^{\prime},B^{\prime}$ instead of $\epsilon,B$).
Of course, finally optimization over $p^{\left(n\right)}$ has to
be performed, which completes the group flow given by Eq. \eqref{final rec}.

The phase transition in the Ising model occurs at zero magnetic field,
so from now we set $B=0$. A necessary condition for optimal $p^{\left(n\right)}$
can be written as $\partial f_{n}/\partial p^{\left(n\right)}=0$,
which (using Eqs. \eqref{f-2} and \eqref{rec2}) turns into:

\begin{equation}
\epsilon=-\frac{1}{\beta}\frac{\left(1-2p^{\left(n\right)}\right)g^{\prime}\left(p^{\left(n\right)}\right)}{4z\left\langle s_{i}^{\left(n\right)}s_{j}^{\left(n\right)}\right\rangle },\label{eps}
\end{equation}
where $g^{\prime}$ denotes a derivative of function $g$. Similarly,
optimization over $p^{\left(n-1\right)}$ leads to the following condition:

\begin{equation}
\epsilon^{\prime}=-\frac{1}{\beta}\frac{\left(1-2p^{\left(n-1\right)}\right)g^{\prime}\left(p^{\left(n-1\right)}\right)}{4z\left\langle s_{i}^{\left(n-1\right)}s_{j}^{\left(n-1\right)}\right\rangle }\label{eps prime}
\end{equation}
and analogous relations hold for parameters $p$ with lower indices.
Dividing Eq. \eqref{eps prime} by Eq. \eqref{eps} and using Eq.
\eqref{final rec} with \eqref{rec2} leads to:

{\small
\begin{align}
 & \left(1-2p^{\left(m\right)}\right)g^{\prime}\left(p^{\left(m\right)}\right)\nonumber \\
 & =\frac{1}{4}\left(1+\frac{1}{\left\langle s_{i}^{\left(m-1\right)}s_{j}^{\left(m-1\right)}\right\rangle }\right)\left(1-2p^{\left(m-1\right)}\right)g^{\prime}\left(p^{\left(m-1\right)}\right),\label{propag}
\end{align}
}for $m=2$. However, as mentioned before, Eq. \eqref{propag} holds
for any $2\leq m\leq n$. If it is satisfied for all these $m$ and
Eq. \eqref{eps} is also satisfied, then $f_{n}/\partial p^{\left(m\right)}=0$
for all $m=1,\cdots,n$. Equation \eqref{propag} can be perceived
as a nonlinear recursion equation for $p^{\left(m\right)}$. Values
of $\left\langle s_{i}^{\left(m\right)}s_{j}^{\left(m\right)}\right\rangle $
have to be simultaneously calculated from \eqref{rec2}. Starting
point can be chosen arbitrarily and terminated at some also arbitrary
index $n$. $\epsilon$ given by Eq. \eqref{eps} will then define
the Hamiltonian which distribution got actually approximated. In other
words, this method defines a reversed renormalization group flow,
but not for parameters of the Hamiltonian, but rather variational
parameters $p^{\left(m\right)}$. Hamiltonian parameters can be read
out by means of Eq. \eqref{eps} (here it is just $\epsilon$).

\begin{figure}
\centering{}\includegraphics[scale=0.4]{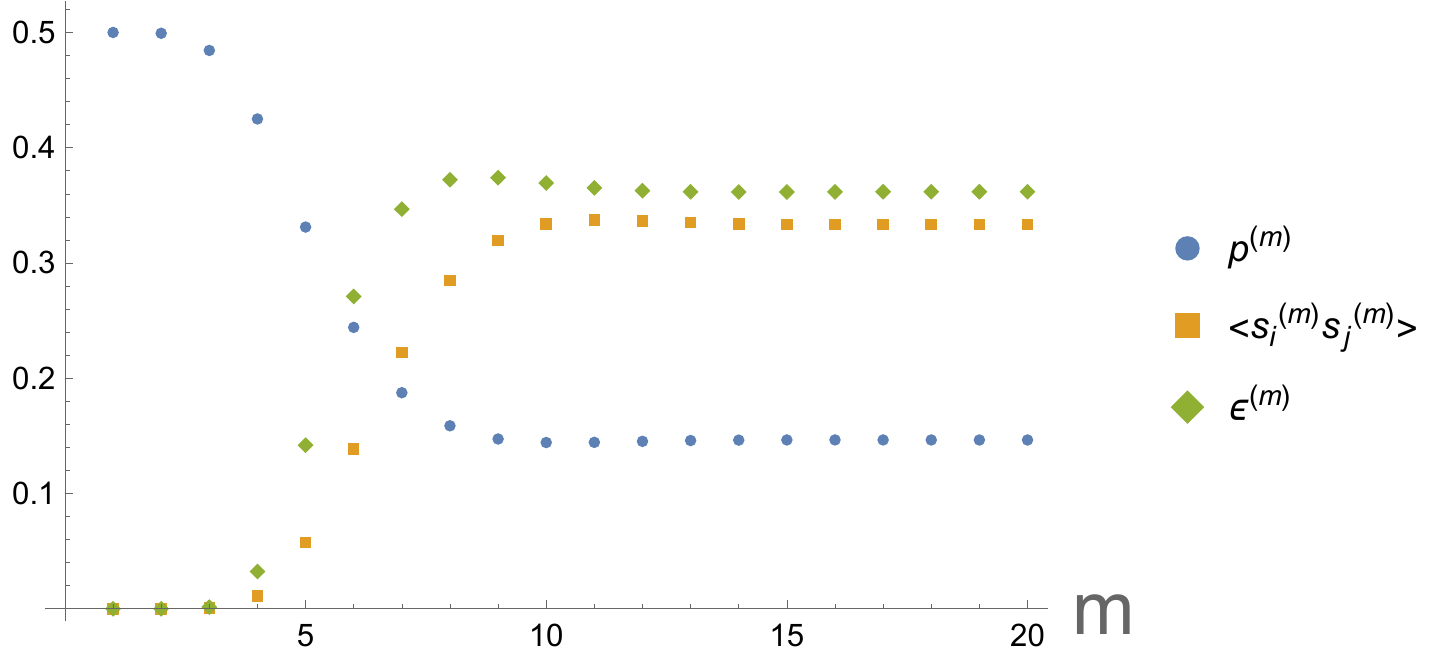}\caption{Evolution of parameters under a reversed renormalization group flow
dictated by the variational method.\label{fig:Evolution-of-parameters}}
\end{figure}

Figure \ref{fig:Evolution-of-parameters} shows an example of the
discussed process. Initial values were $p^{\left(0\right)}=0.5-10^{-4}$
and $\left\langle s_{i}^{\left(0\right)}s_{j}^{\left(0\right)}\right\rangle =0$.
It is seen that the parameters converge to their limits, which are
interpreted as critical values. They can be found analytically by
looking for a fixed point of recursion Eqs. \eqref{rec2} and \eqref{propag}.
From the latter, substituting $p^{\left(m\right)}=p^{\left(m-1\right)}$
and cancelling terms involving $p$, we get $\left\langle s_{i}s_{j}\right\rangle _{c}=1/3$
(subscript $c$ refers to criticality). Taking Eq. \eqref{rec2} and
setting the spin-spin correlation to $1/3$ yields $2\left(1-2p_{c}\right)^{2}=1$,
so $p_{c}=\left(2-\sqrt{2}\right)/4\approx0.146$. Then, Eq. \eqref{eps}
gives the critical value of $\epsilon_{c}$:

\begin{equation}
\epsilon_{c}=-\frac{1}{\beta}\frac{3g^{\prime}\left(p_{c}\right)}{16\sqrt{2}}.
\end{equation}

Setting $\beta=0.5$ (to match the convention from \citep{Binney})
gives $\epsilon_{c}=0.362$. Equivalently, $\epsilon$ can be set
to $0.5$ (as in section \ref{sec:Percolation-based-drawing-for}),
which gives $\beta_{c}=0.362$. The exact value is $0.4407$ \citep{Binney},
so the approximation is rather crude. The drawing scheme used here
is a minimal one exhibiting fractal structure. It improves significantly
the mean-field result $\beta_{c}=0.25$, but most importantly, it
provides critical exponents from the framework of renormalization
group.

\begin{figure}
\centering{}\includegraphics[scale=0.42]{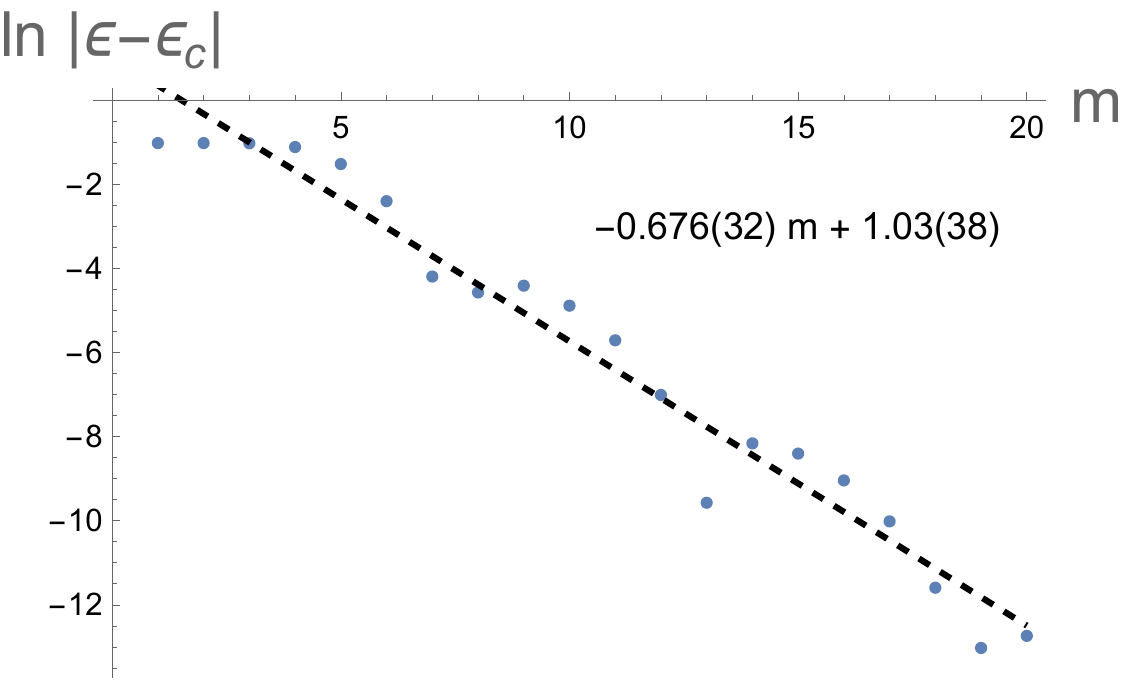}\caption{Determination of $\nu$ as $\ln b/\ln\lambda_{T}$. The denominator
of this expression equals minus the slope.\label{fig:Determination-of-}}
\end{figure}

Critical exponent $\nu=\ln b/\ln\lambda_{T}$ \citep{Binney}, where
$\left|\epsilon^{\left(m\right)}-\epsilon_{c}\right|$ falls like
$1/\lambda_{T}^{m}$ for large $m$. Thus, obtaining $\ln\lambda_{T}$
from a sequence $\ln\left|\epsilon^{\left(m\right)}-\epsilon_{c}\right|$
can be done by a simple linear fitting, which is depicted in Fig.
\ref{fig:Determination-of-}. This leads to $\nu=1.025(49)$. The
fitting and standard errors were obtained using the ``ParameterTable''
option of the ``NonlinearModelFit'' function in Mathematica.

A more sophisticated approach to the same task can be realized by
examining asymptotic behavior of both $p^{\left(m\right)}$ and $\left\langle s_{i}^{\left(m\right)}s_{j}^{\left(m\right)}\right\rangle $
as $m\rightarrow\infty$. Equations \eqref{propag} and \eqref{rec2}
define a transformation $\left(x,y\right)\rightarrow R\left(x,y\right)$,
which maps $p^{\left(m\right)}$ and $\left\langle s_{i}^{\left(m\right)}s_{j}^{\left(m\right)}\right\rangle $
(in the role of $x,y$) to their new values. If a fixed point $\left(x_{c},y_{c}\right)$
is identified, then for $\left(x,y\right)$ close to $\left(x_{c},y_{c}\right)$
it suffices to expand $R$ in a first-order Taylor series (around
the fixed point) to investigate the dynamics of $\left(x,y\right)$
(under iterating $R$). Let $J$ be a Jacobian of $R$ (at the fixed
point), so that:

\begin{equation}
R\left(x_{c}+\Delta x,y_{c}+\Delta y\right)\cong R\left(x_{c},y_{c}\right)+J\begin{pmatrix}\Delta x\\
\Delta y
\end{pmatrix}.
\end{equation}

Then:

\begin{equation}
\begin{pmatrix}\Delta x^{\left(m+1\right)}\\
\Delta y^{\left(m+1\right)}
\end{pmatrix}\cong J\begin{pmatrix}\Delta x^{\left(m\right)}\\
\Delta y^{\left(m\right)}
\end{pmatrix}.
\end{equation}
Thus eigenvalues of $J$ dictate the asymptotic behavior of $\left(\Delta x,\Delta y\right)$.
The greater (with respect to its norm) eigenvalue governs the decay
of the norm of $\left(\Delta x,\Delta y\right)$. Then $\epsilon$
given by Eq. \eqref{eps} can also be expanded in a first-order Taylor
series around the critical values:

\begin{equation}
\epsilon-\epsilon_{c}\cong\frac{\partial\epsilon}{\partial x}\Delta x+\frac{\partial\epsilon}{\partial y}\Delta y.\label{eps Taylor}
\end{equation}

Numerical values of $J$, $\partial\epsilon/\partial x$ and $\partial\epsilon/\partial y$
are:

\begin{equation}
J=\left(\begin{array}{cc}
1 & 0.216914\\
-1.88562 & -0.159017
\end{array}\right),\label{J}
\end{equation}

\begin{equation}
\left(\frac{\partial\epsilon}{\partial x},\frac{\partial\epsilon}{\partial y}\right)=\left(-3.75317,-1.08549\right).\label{grad eps}
\end{equation}
The eigenvalues of $J$ are $0.420491\pm0.270531\mathrm{i}$, with
norm $0.5$. Symbolic calculation in Mathematica reveals that $1/2$
is an exact value of this norm. Any vector $\left(\Delta x,\Delta y\right)$
can be decomposed into eigenvectors of $J$. Each of these two components
shrinks by a factor of $1/2$ after every iteration. The overall norm
of $\left(\Delta x,\Delta y\right)$ may exhibit some fluctuations
due to phase difference in the components, which are probably visible
in Fig. \ref{fig:Determination-of-}. They are likely to be unrealistic
artifacts of the used drawing scheme, in which all bonds are not actually
on equal footing. For more sophisticated drawing schemes, this effect
should be naturally minimized. Since Eq. \eqref{eps Taylor} links
$\epsilon-\epsilon_{c}$ to the magnitude of $\left(\Delta x,\Delta y\right)$,
$1/\lambda_{T}=1/2$ is obtained. It gives $\nu=1$, which is the
exact value of this critical exponent.

The magnetic eigenvalue $\lambda_{B}$ governing the scaling of $B$
can be obtained very easily from Eq. \eqref{final rec}. Close to
the critical point, $B$ is very small and $p^{\left(n\right)}$ is
very close to $p_{c}$, so:

\begin{equation}
B^{\prime}\cong4\left(1-2p_{c}\right)B,
\end{equation}
which gives $\lambda_{B}=4\left(1-2p_{c}\right)$. Using $\eta=4-2\log\lambda_{B}/\log b$
\citep{Binney} yields $\eta=1$. This is a huge overestimation ($\eta=1/4$
is the exact value), but already from the critical value $\epsilon_{c}$,
the crudeness of the proposed drawing was visible. However, the method
is capable of handling more complex up-normalization and down-normalization
procedures.

As foreshadowed, a comparison of the presented approach with the standard
Kadanoff's variational renormalization group \citep{Kadanoff} will
be given. First of all, the latter is focused on finding the renormalized
form of the Hamiltonian and establishing this way the group flow.
In the method of this paper, multi-stage drawing schemes are optimized
to convey the statistical behavior of the system. Group flow emerges
for the parameters residing in the stages of the drawing procedure.

Hamiltonians and drawing schemes can in principle carry the same information,
but what looks easy in one representation may be heavy in the other,
so the presented method has a potential for handling otherwise cumbersome
cases. Also, it is easy to control whether a proposed drawing scheme
is still tractable (it generally is, as long as it is a sequence of
simple decisions).

Work \citep{Kadanoff} allows approaching the free energy from both
sides (Eq. (5)). Formula (6) provides renormalization giving an upper
bound, while Eq. (7) gives a lower bound. In this paper's approach
only upper bound is realized. Its perspective can be partially applied
to Kadanoff's Eq. (6). Projection function $S\left(\mu,\sigma\right)$
can be viewed as down-normalization and $H_{0}\left(\mu,\sigma\right)$
as up-normalization. However, even with this prescription, the methods
differ fundamentally in previously mentioned aspects.

\section{Conclusion}

A short summary of qualitative features of each presented drawing
scheme will be given. The $k$-space mean-field indicates a lower
critical dimensions of $3$ (instead of $2$), but its advantage over
the real-space mean-field is significance of taking the thermodynamic
limit to obtain a phase transition. The results have a similar structure
to those of the spherical model.

The next procedure involving site-independent drawing followed by
a linear transformation $T$ effectively includes two-point correlations
as variable parameters. Therefore, long-range correlations can be
accounted for more precisely. It manifests a phase transition in two
dimensions, but behavior of the system at the critical point is not
captured correctly. This can be attributed to the lack of fractal
structure present in the drawing scheme, which is expected at the
critical point.

Section about the percolation-based method provides a general double
drawing scheme approach (i. e. involving both $P_{1}$ and $P_{2}$),
which may facilitate new creative designs of state drawing for various
models. Of course, including $P_{2}$ comes at a cost. A larger system
is modeled (in the considered case bonds plus spins), so the space
of all configurations is even larger. However, presence of $P_{2}$
is needed, because it resolves difficulties with sums under the logarithm
in the entropy term. Knowledge of percolation was a substantial part
of the presented strategy. It was gained by performing a Monte-Carlo
algorithm \citep{Newman}, which raises a question why not to simulate
the Ising model directly. Of course, the effort in this specific instant
would be less. However, the presented variational framework allows
to map many different models onto percolation. This mapping does not
have to be exact, so is more flexible and universal. Presented drawing
procedure exhibits in some sense fractal-structure, due to the presence
of percolation model in it. However, indication of critical exponents
would require knowledge of non-analytic behavior of percolation-related
functions as $P\left(p\right)$ and $M\left(p\right)$. Critical temperature
can be determined without it, just by knowing the percolation threshold
and numerical values of $P\left(p\right)$ and $M\left(p\right)$.

Iterative fractal-like drawing schemes use the concept of double drawing
and are able to give results in the style of the renormalization group.
In standard calculations of the latter method two steps are needed
to obtain the group flow: renormalization scheme (e. g. decimation
or majority rule) and some approximation to write the renormalized
Hamiltonian in the constrained functional form. In the presented method,
establishing the group flow requires proposing up-normalization and
down-normalization schemes (involving adjustable parameters) and minimizing
the free energy over the parameters. All approximations are of variational
nature (i. e. they come from the limited generality of the drawing
procedures).

The main findings of the paper are developing the method of handling
intricate drawing procedures (double drawing scheme) and identifying
the need of fractality in these schemes to grasp the critical behavior.
Presented ideas can be extended for quantum mechanical systems by
developing an idea of drawing quantum states.

\end{document}